\documentclass[twocolumn]{aa}

\usepackage[T1]{fontenc}
\usepackage[utf8]{inputenc}

\usepackage{graphicx}
\usepackage{natbib}
\usepackage{ulem}
\usepackage{textcomp}
\usepackage{gensymb}
\usepackage{longtable}
\usepackage{threeparttable}
\usepackage{multicol}
\usepackage{multirow}
\usepackage{textgreek}
\usepackage{float}
\usepackage{todonotes}
\usepackage[Symbol]{upgreek}
\usepackage{amsmath}
\usepackage{etoolbox}
\usepackage{txfonts}
\usepackage{url}
\usepackage{xcolor}
\usepackage{xfrac}

\usepackage[most]{tcolorbox}
\usepackage{hyperref}
\usepackage{xcolor}
\hypersetup{
  colorlinks   = true, 
  urlcolor     = black, 
  linkcolor    = blue, 
  citecolor   = blue 
}
\usepackage[all]{hypcap} 

\begin{document} 
	\title{Physical insights from the spectrum of the radio halo in MACS\,J0717.5$+$3745}
	\titlerunning{Physical insights from the spectrum of the radio halo in MACS\,J0717.5$+$3745}
	\authorrunning{Rajpurohit et al.}

\author{K. Rajpurohit\inst{1,2,3},  G. Brunetti\inst{2}, A. Bonafede\inst{1,2,4}, R. J. van Weeren\inst{5}, A. Botteon\inst{5}, F. Vazza\inst{1,2,4}, M. Hoeft\inst{3}, C. J. Riseley\inst{1,2}, E. Bonnassieux\inst{1,2},  M. Brienza\inst{1,2}, W. R. Forman\inst{6}, H. J. A. R\"ottgering\inst{5}, A. S. Rajpurohit\inst{7}, N. Locatelli\inst{1,2}, T. W. Shimwell\inst{8,5}, R. Cassano\inst{2}, G. Di Gennaro\inst{5}, M. Br\"uggen\inst{4}, D. Wittor\inst{4}, A. Drabent\inst{3}, and A. Ignesti\inst{1,2}  }
\institute{Dipartimento di Fisica e Astronomia, Universit\'at di Bologna, via P. Gobetti 93/2, 40129, Bologna, Italy\\
 {\email{kamlesh.rajpurohit@unibo.it}}
\and
INAF-Istituto di Radio Astronomia, Via Gobetti 101, 40129 Bologna, Italy
\and
Th\"uringer Landessternwarte, Sternwarte 5, 07778 Tautenburg, Germany
\and
Hamburger Sternwarte, Universit\"at Hamburg, Gojenbergsweg 112, 21029, Hamburg, Germany
\and
Leiden Observatory, Leiden University, P.O. Box 9513, 2300 RA Leiden, The Netherlands
\and
Harvard-Smithsonian Center for Astrophysics, 60 Garden Street, Cambridge, MA 02138, USA
\and
Astronomy \& Astrophysics Division, Physical Research Laboratory, Ahmedabad 380009, India
\and
ASTRON, the Netherlands Institute for Radio Astronomy, Postbus 2, 7990 AA Dwingeloo, The Netherlands
}

 %%%%%%%%%%%%%%%%%%%%%%%%%%%%%%%%%%%%%%%%%%%%%%%%%%%%%%%%%%%%%%%%%%%%%%%%%%%%%%%%%%%
 % Abstract
 %%%%%%%%%%%%%%%%%%%%%%%%%%%%%%%%%%%%%%%%%%%%%%%%%%%%%%%%%%%%%%%%%%%%%%%%%%%%%%%%%%%

     \abstract
  {We present new LOFAR observations of the massive merging galaxy cluster MACS\,J0717.5$+$3745, located at a redshift of $0.5458$. The cluster hosts the most powerful radio halo known to date. These new observations, in combination with published uGMRT (300$-$850\,MHz) and VLA (1$-$6.5\,GHz) data, reveal that the halo is more extended than previously thought, with a largest linear size of $\sim2.2\,\rm Mpc$, making it one of the largest known halos. The halo shows a steep spectrum ($\alpha_{144\,\text{MHz}}^{1.5\,\text{GHz}}\sim-1.4$) and a steepening ($\alpha_{1.5\,\text{GHz}}^{5.5\,\text{GHz}}\sim-1.9$) above 1.5\,GHz. We find  a strong scattering in spectral index maps on scales of 50$-$100\,kpc. We suggest that such a strong scattering may be a consequence of the regime where inverse Compton dominates the energy losses of electrons. The spectral index becomes steeper and shows an increased curvature in the outermost regions of the halo. We combined the radio data with \textit{Chandra} observations to investigate the connection between the thermal and non-thermal components of the intracluster medium (ICM). Despite a significant substructure in the halo emission, the radio brightness correlates strongly with the X-ray brightness at all observed frequencies. The radio-versus-X-ray brightness correlation slope steepens at a higher radio frequency (from $b_{144\,\text{MHz}}=0.67\pm0.05$ to $b_{3.0\,\text{GHz}}=0.98\pm0.09$) and the spectral index shows a significant anti correlation with the X-ray brightness. Both pieces of evidence further support a spectral steepening in the external regions. The compelling evidence for a steep spectral index, the existence of a spectral break above 1.5\,GHz, and the dependence of radio and X-ray surface brightness correlation on frequency are interpreted in the context of turbulent reacceleration models. Under this scenario, our results allowed us to constrain that the turbulent kinetic pressure of the ICM is up to 10\%.}
   

  \maketitle

%%%%%%%%%%%%%%%%%%%%%%%%%%%%%%%%%%%%%%%%%%%%%%%%%%%%%%%%%%%%%%%%%%%%%%%%%%%%%%%%%%%%
%  Introduction
%%%%%%%%%%%%%%%%%%%%%%%%%%%%%%%%%%%%%%%%%%%%%%%%%%%%%%%%%%%%%%%%%%%%%%%%%%%%
\section{Introduction}
 \label{sec:intro}
  
Galaxy clusters host non-thermal components in the form of $\upmu \rm G$ magnetic fields and relativistic particles mixed with the thermal intracluster medium (ICM). In many merging clusters, these non-thermal components manifest as highly-extended synchrotron sources, broadly classified as radio halos and radio relics. The emission in these sources is not associated with individual galaxies in the  cluster, hinting at a large-scale (re)acceleration mechanism powering the radio emitting electrons \citep[see][for a recent review]{vanWeeren2019}. The radio spectra of such sources are steep{\footnote{$S_{\nu}\propto\nu^{\alpha}$, with spectral index $\alpha$}} ($\alpha \leq-1$). 

Both halos and relics are typically associated with merging clusters \citep[e.g.,][]{Buote2001,Giacintucci2008,Cassano2010,Finoguenov2010,vanWeeren2011}, suggesting that cluster mergers play a major role in their formation. However, despite progress in understanding of radio halos and relics, the exact formation mechanism is not fully understood \citep[][for a review]{Brunetti2014}.

Our paper focuses on the radio halo in the galaxy cluster MACS\,J0717.5$+$3745. Giant radio halos are megaparsec scale sources that are located in the center of a cluster and are usually unpolarized, regardless of the frequencies at which they have been observed. The radio emission from halos typically follows the X-ray emission \citep{Govoni2001b,Pearce2017,Rajpurohit2018,Botteon2020}. Statistical studies of radio halos reveal various correlations between the radio power at 1.4\,GHz and thermal properties of the cluster, such as mass and X-ray luminosity \citep[e.g.,][]{Cassano2013}, supporting a tight connection between the the thermal and non-thermal components in the ICM. 

%####################################
% obs properties
%####################################
\setlength{\tabcolsep}{8.0pt}   
\begin{table*}[!htbp]
\caption{Summary of observations }
\centering
\label{Table 2}
 \begin{threeparttable} 
\begin{tabular}{c c c c c c c c r}
\hline\hline
Band & Configuration  &Frequency range &Central frequency &Channel Width & On Source Time & LAS$^{_\ddagger}$ \\ % table heading
&&&&&hour&\\
\hline
LOFAR&  HBA &120-168 MHz & 144 MHz &12.2 kHz&16&$3.8\degree$ \\
uGMRT $^{\dagger}$ &band 3 &330-500 MHz& 400 MHz&97 kHz&6 &1920$\arcsec$ \\
uGMRT $^{\dagger}$& band 4 &550-850 MHz& 700 MHz&97 kHz&8&1020$\arcsec$\\
VLA L-band $^{\dagger}$&  ABCD&1-2 GHz & 1.5 GHz&1 MHz&15&970$\arcsec$ \\
VLA S-band $^{\dagger}$&  ABCD&2-4 GHz&3.0 GHz &2 MHz&13&490$\arcsec$ \\
VLA C-band $^{\dagger}$& BCD&4.5-6.5 GHz& 5.5 GHz&2 MHz&12&240$\arcsec$\\
 \hline
\\
\end{tabular}
\begin{tablenotes}[flushleft]
  \footnotesize
   \vspace{-0.2cm}
   \item\textbf{Notes.}  Full Stokes polarization information were recorded for VLA, LOFAR and uGMRT band 4 data; $^{_\dagger}$ For data reduction steps, we refer to \cite{vanWeeren2016b,vanWeeren2017b,Rajpurohit2020c}; $^{_\ddagger}$Largest angular scale (LAS) that can be recovered by mentioned observations.  
       \end{tablenotes}
    \end{threeparttable} 
\label{Tabel:obs}
\end{table*}

The currently favored scenario for the formation of radio halos involves the re-acceleration of cosmic ray electrons (CRe) to higher energies by  turbulence induced during mergers \citep[e.g.,][]{Brunetti2001,Petrosian2001,Brunetti2007,Miniati2015,Brunetti2016}. The generation of secondary particles via hadronic collisions in the ICM provides an additional mechanism \citep{Dennison1980,Blasi1999}. However, the current gamma ray limits imply that the energy budget of cosmic rays is too small to explain radio halos \citep{Jeltema2011,Brunetti2012,Ackermann2014,Ackermann2016}. Nevertheless, secondary particles may still contribute a relevant part of the pool of electrons that are reaccelerated in the reacceleration scenario \citep[e.g., ][]{Brunetti2017,Pinzke2017}.

Radio halos have been observed with a range of spectral indices, mainly between $-1.1$ to $-1.4$ \citep{vanWeeren2019}. The discovery of ultra-steep spectrum radio halos, that is $\alpha<-1.5$ \citep{Brunetti2008,Bonafede2012,Giacintucci2013,Wilber2018} has also provided support for the turbulent reacceleration model, connecting steeper spectral indices with more inefficient acceleration stages. However, despite much progress in both theory and observations, our understanding of the complex chain of physical processes that connect mergers with the generation and evolution of turbulence and the acceleration of particles remains limited.

%%%%%%%%%%%%%%%%%%%%%%%%%%%%%%%%%%%%%%%%%%%%%%%%%%%%%%%%%%%%%%%%%%%%%%%%%%%%%%%%%%%%
% Target
%%%%%%%%%%%%%%%%%%%%%%%%%%%%%%%%%%%%%%%%%%%%%%%%%%%%%%%%%%%%%%%%%%%%%%%%%%%%%%%%%%%% 

The galaxy cluster MACS\,J0717.5$+$3745, located at a redshift of $z = 0.5458$, is one of the most complex dynamically disturbed systems \citep[e.g.,][]{Ebeling2001,Edge2003}. Optical and X-ray observations of MACS\,J0717.5$+$3745 revealed that it consists of at least four components belonging to different merging subclusters \citep{Limousin2016,vanWeeren2017b}. Recently, \cite{Jauzac2018} reported that the cluster is dominated by 9 subgroup-like structures. The X-ray luminosity of the cluster is $L_{\rm x, \rm 0.1-2.4\,keV}=2.4\times10^{45}\,\rm erg\,s^{-1}$. It is one of the hottest clusters known currently with an overall ICM temperature of $12.2\pm0.4\rm\,keV$ \citep{Ebeling2007,vanWeeren2017b}. The mass of the cluster estimated from the Sunyaev-Zeldovich (SZ)  Planck signal is $\rm M_{500}=(11.487\pm 0.54)\times10^{14} \rm\,M_{\odot}$ \citep{Planck2016}.

\begin{figure*}[!thbp]
    \centering
    \includegraphics[width=0.49\textwidth]{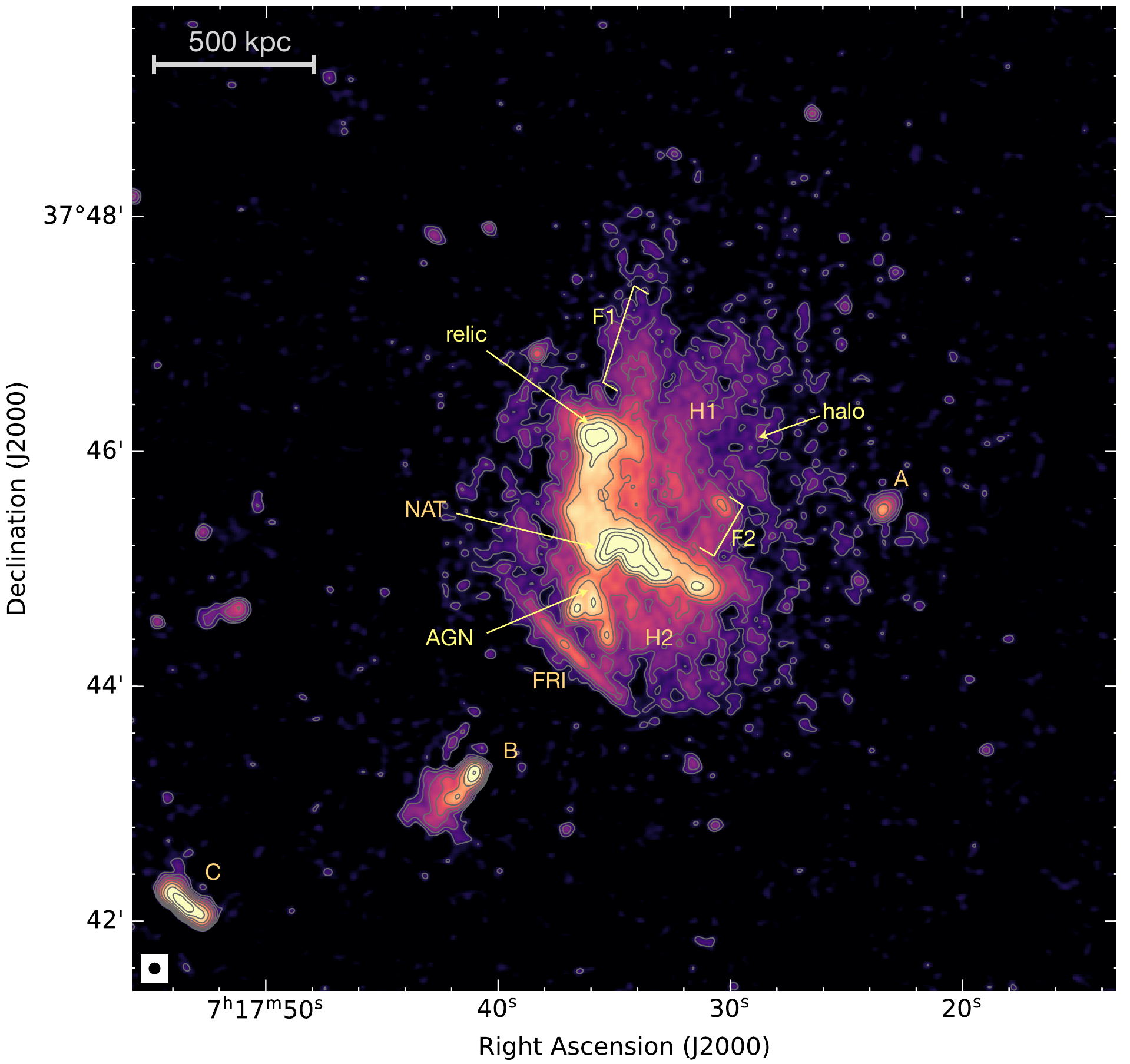}
      \includegraphics[width=0.49\textwidth]{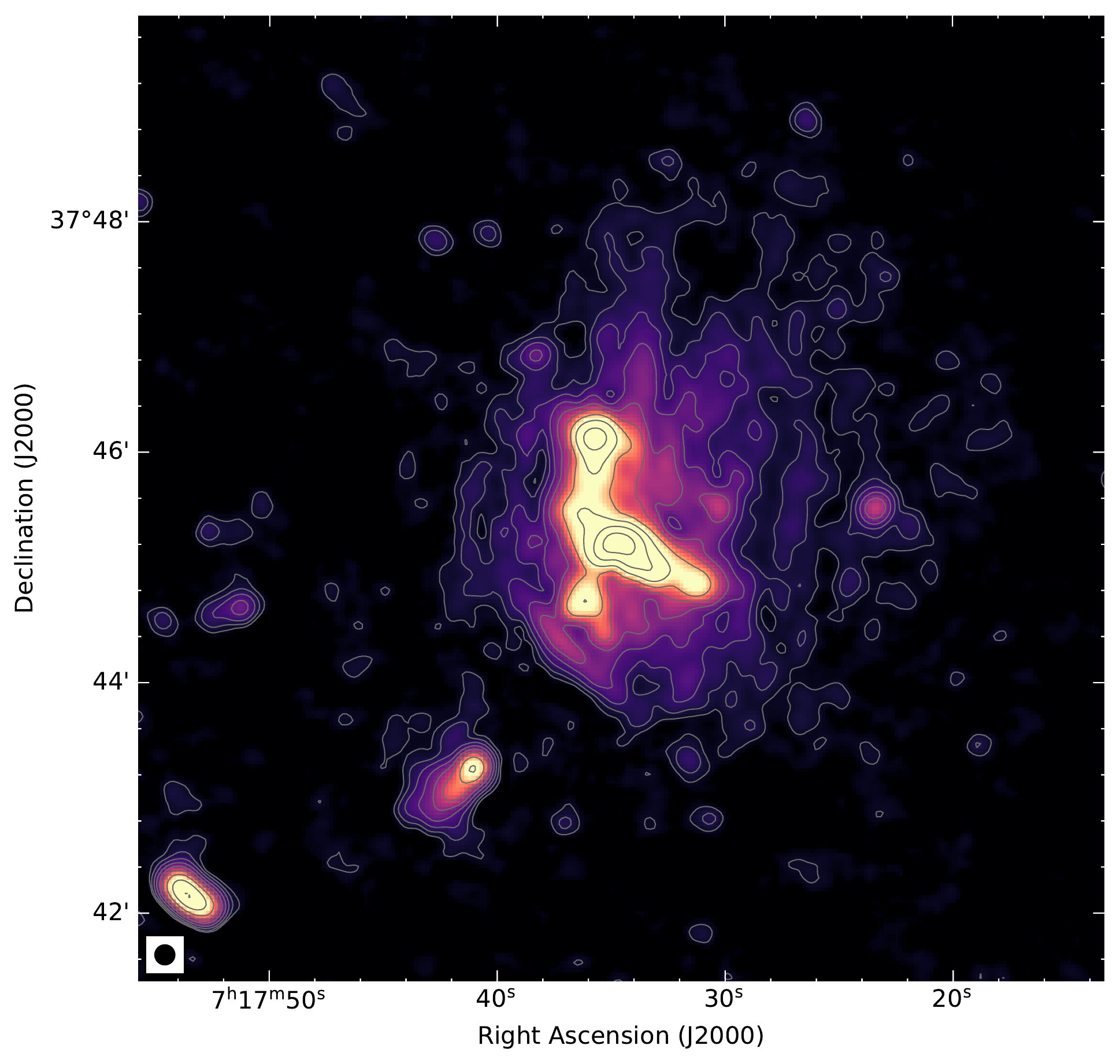}       \\
     \vspace{0.3cm}
          \includegraphics[width=0.49\textwidth]{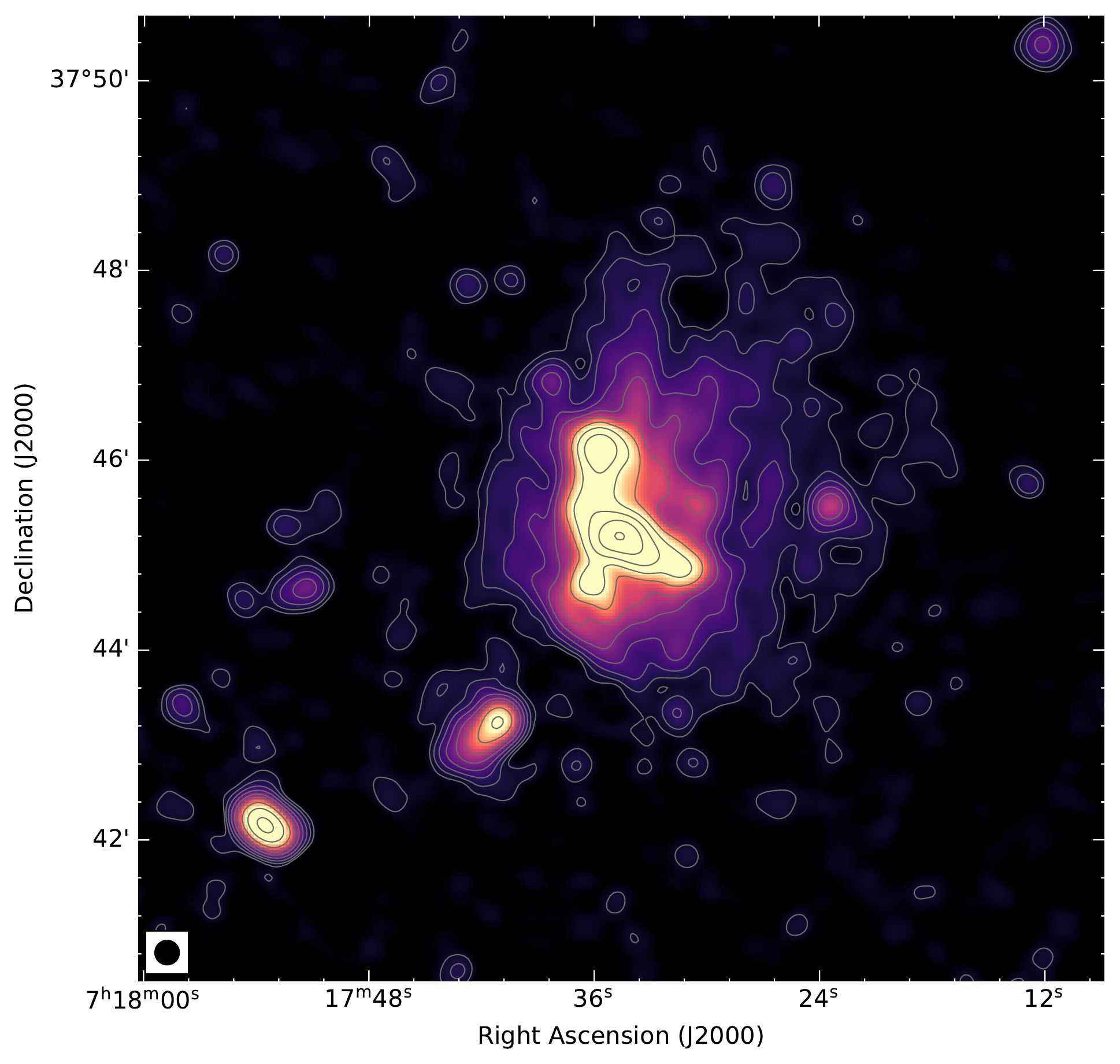}
      \includegraphics[width=0.49\textwidth]{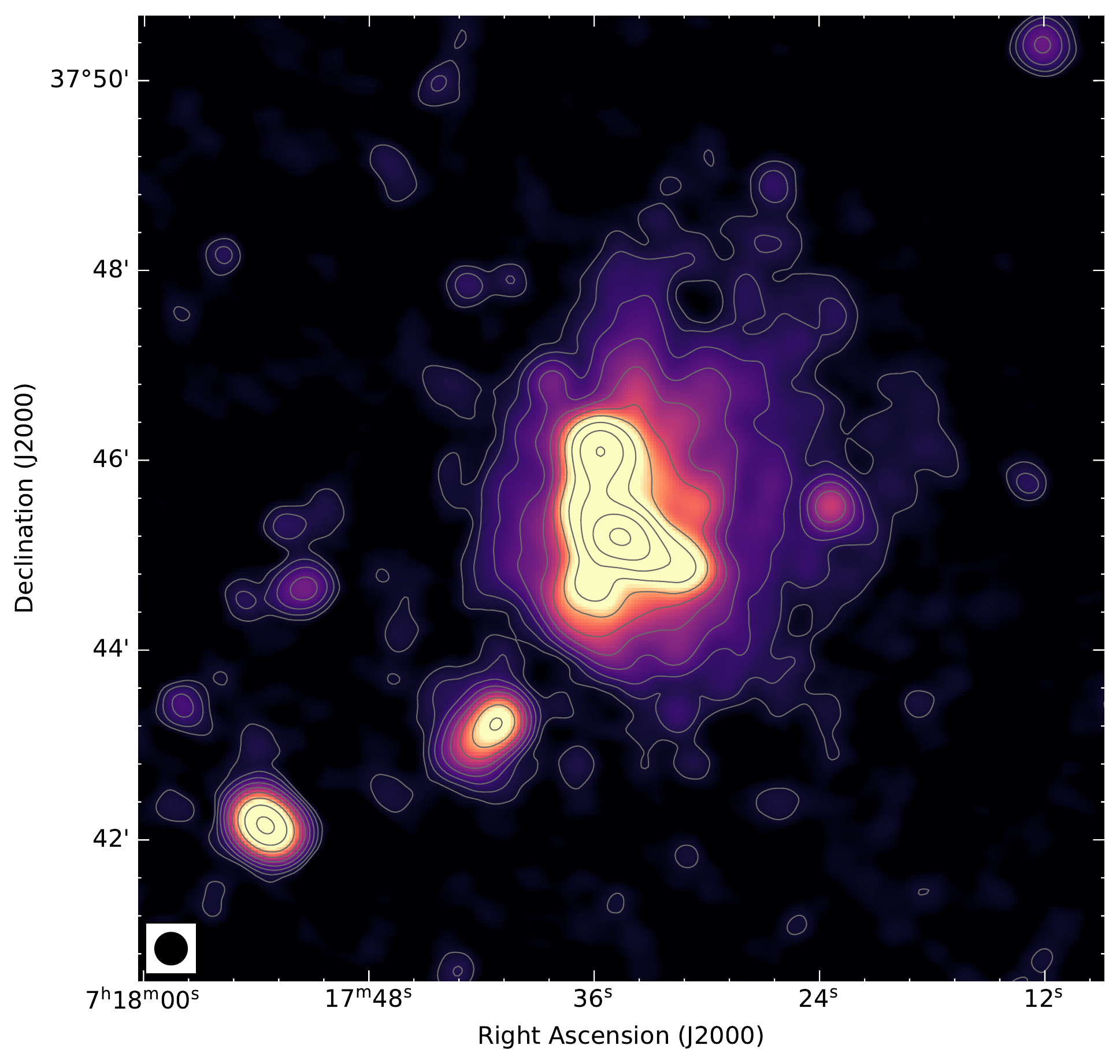}
    \vspace{-0.3cm}
    \caption{Deep LOFAR 144\,MHz images of the galaxy cluster MACS\,J0717+3745 at different resolutions ($5\arcsec$, top left; $10\arcsec$, top right; $15\arcsec$, bottom left; $20\arcsec$, bottom right). The known diffuse emission sources---the large-scale halo and the chair-shaped relic---are well recovered by LOFAR at 144\,MHz. The halo consists of the northern (H1) and southern (H2) regions. Contour levels are drawn at $[1,2,4,8,\dots]\,\times\,4.0\,\sigma_{{\rm{ rms}}}$, where $\sigma_{{\rm{rms,5\arcsec}}}=74\,\rm \upmu Jy\,beam
   ^{-1}$, $\sigma_{{\rm{rms,10\arcsec}}}=90\,\rm \upmu Jy\,beam
   ^{-1}$, $\sigma_{{\rm{rms,15\arcsec}}}=112\,\rm \upmu Jy\,beam
   ^{-1}$, and $\sigma_{{\rm{rms,25\arcsec}}}=133\,\rm \upmu Jy\,beam
   ^{-1}$. In these images there is no region below $-4\sigma_{{\rm rms}}$. The images are created with {\tt Briggs} weighting and ${\tt robust}=-0.20$. The beam sizes are indicated in the bottom left corner of the each image.}
      \label{fig1}
\end{figure*}

\begin{figure*}[!thbp]
    \centering
    \includegraphics[width=1.0\textwidth]{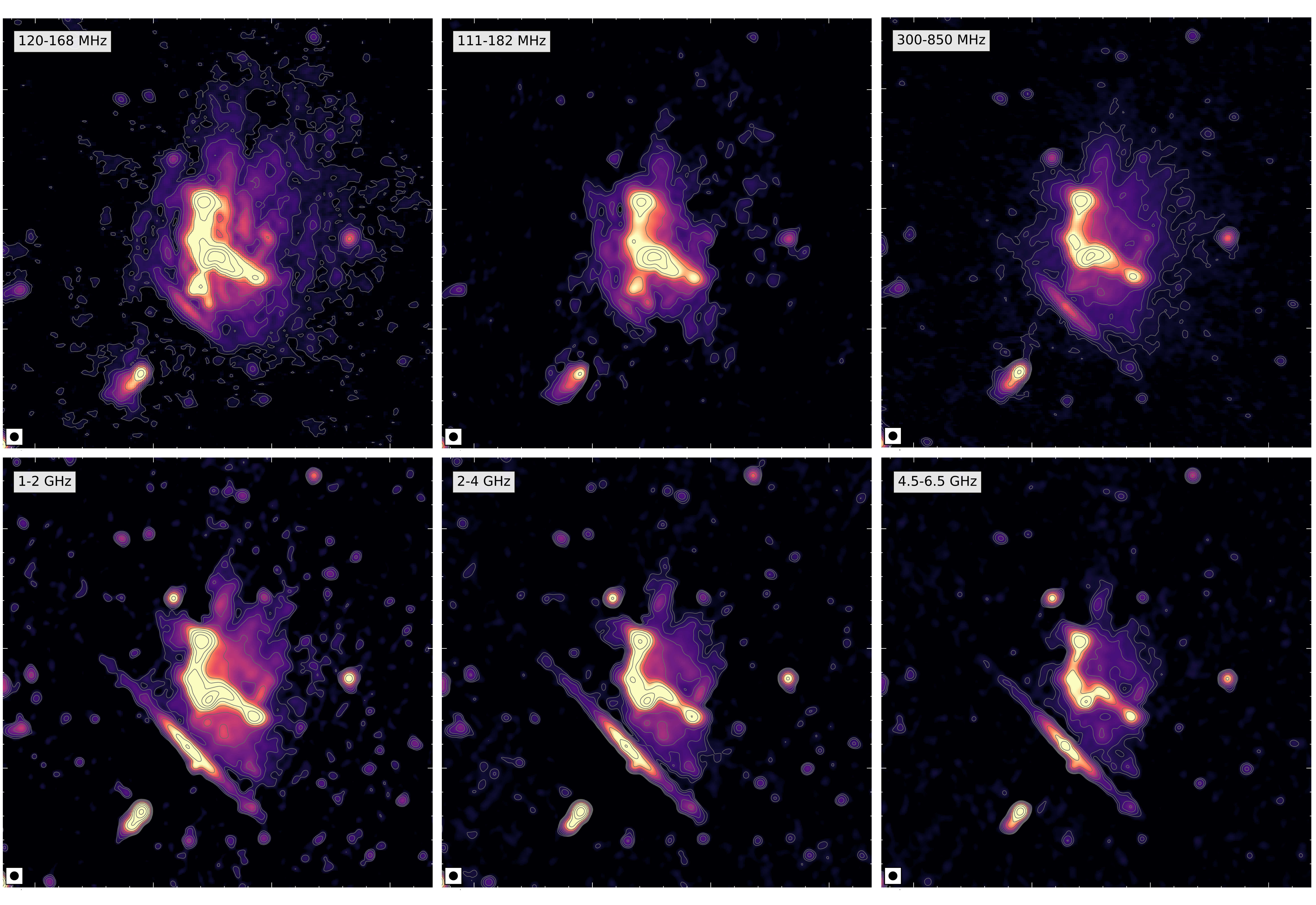}
    \vspace{-0.7cm}
    \caption{Comparison of the halo from 144\,MHz to 5.5\,GHz. All images are at created at a common $8\arcsec$ resolution. The top-middle panel image (111-182 MHz) is adopted from \cite{Bonafede2018}. The colors in all images were scaled manually. It is evident that the halo emission is more extended towards low frequencies. Contour levels are drawn at $[1,2,4,8,\dots]\,\times\,3.0\,\sigma_{{\rm{ rms}}}$, where $\sigma_{{\rm{rms,\rm 144\,MHz}}}=82\,\rm \upmu Jy\,beam
   ^{-1}$, $\sigma_{{\rm{rms,\,147\,MHz}}}=130\,\rm \upmu Jy\,beam
   ^{-1}$, $\sigma_{{\rm{rms,\, 575\,MHz}}}=23\,\rm \upmu Jy\,beam
   ^{-1}$, $\sigma_{{\rm{rms,\, 1.5\,GHz}}}=7\,\rm \upmu Jy\,beam
   ^{-1}$, $\sigma_{{\rm{rms,\, 3.4\,GHz}}}=3\,\rm \upmu Jy\,beam
   ^{-1}$, and $\sigma_{{\rm{rms,\, 5.5\,GHz}}}=3.2\,\rm \upmu Jy\,beam
   ^{-1}$ The beam sizes are indicated in the bottom left corner of the each image.}
      \label{compare}
\end{figure*}

MACS\,J0717.5$+$3745 hosts one of the most complex radio halos, extending over an area of at least 1.6\,Mpc \citep{Bonafede2009b,vanWeeren2009,PandeyPommier2013,vanWeeren2017b,Bonafede2018}. The 1.4\,GHz radio luminosity of the halo is the highest known for any cluster, in agreement with the cluster's large mass and high global temperature \citep[e.g.,][]{Cassano2013}. The detection of polarized emission  was claimed by \cite{Bonafede2009b}.

The cluster was observed by the LOw-Frequency ARray \citep[LOFAR;][]{Haarlem2013} High Band Antenna (HBA) by \cite{Bonafede2018}. These observations revealed a large 1.9\,Mpc radio arc to the northwest of the main relic and a radio bridge that connects the cluster to a head-tail radio galaxy located along a filament of galaxies falling into the main cluster. 

Taking advantage of improvements in data processing pipelines, as well as data from the LOFAR (120-168\,MHz) Two-metre Sky Survey \citep[LoTSS;][]{Shimwell2017}, in this paper we describe the results of new observations of the galaxy cluster MACS\,J0717.5$+$3745. These observations are complemented by  VLA L-band (1-2\,GHz), S-band (2-4\,GHz), C-band (4.5-6.5\,GHz) observations, as well as uGMRT Band 3 (300$-$500\,MHz) and Band 4 (550$-$850\,MHz) observations. The outline of this paper is as follows: In Sect.\,\ref{obs}, we describe the observations and data reduction. The total power images are presented in Sect.\,\ref{results}. This is followed by a detailed analysis and discussion in Sect.\,\ref{discussion} and Sect.\,\ref{discussion1}, respectively. We summarize our main findings in Sect.\,\ref{summary}. 

Throughout this paper we assume a $\Lambda$CDM cosmology with $H_{\rm{ 0}}=70$ km s$^{-1}$\,Mpc$^{-1}$, $\Omega_{\rm{ m}}=0.3$, and $\Omega_{\Lambda}=0.7$. At the cluster's redshift, $1\arcsec$ corresponds to a physical scale of 6.4\,kpc. All output images are in the J2000 coordinate system and are corrected for primary beam attenuation.

%%%%%%%%%%%%%%%%%%%%%%%%%%%%%%%%%%%%%%%%%%%%%%%%%%%%%%%%%%%%%%%%%%%%%%%%%%%%%%%%%%%
%Observation and data reduction}
%%%%%%%%%%%%%%%%%%%%%%%%%%%%%%%%%%%%%%%%%%%%%%%%%%%%%%%%%%%%%%%%%%%%%%%%%%%%%%%%%%%%
\section{Observation and data reduction}
\label{obs}

The first LOFAR HBA observations of MACS\,J0717.5$+$3745 were presented by \citet{Bonafede2018}. In that work, the authors used 5~hr observations carried out in Cycle 0, and the data were calibrated using using {\tt FACTOR} \citep{vanWeeren2016c}. Here, we use newer data from LoTSS, see Table\,\ref{Tabel:obs} for observational details. 

MACS\,J0717.5$+$3745 was covered  by two LoTSS pointings (P110+39 and P109+37) on 2019 December 5 and 2020 January 21. The phase center were offset by $\sim1.9\degree$ and $<1\degree$ from the MACS\,J0717.5$+$3745 cluster center for P110+39 and P109+37, respectively.  Since the LOFAR field of view is  about $5\degree$ at 120-168\,MHz, this phase offset do not affect our observations.  Each LoTSS pointing consists of an 8~hr observation book-ended by 10~min scans of the flux density calibrator using HBA stations in the \verb|HBA_DUAL_INNER| mode, and is processed with the standard Surveys Key Science Project pipeline\footnote{\url{https://github.com/mhardcastle/ddf-pipeline}} (see \citealt{Shimwell2019,Tasse2020}). This pipeline aims to correct for direction-independent and direction-dependent effects using \verb|PREFACTOR| \citep{vanWeeren2016c, Williams2016, deGasperin2019} and \verb|KillMS| \citep{Tasse2014a, Tasse2014b, Smirnov2015}, and to produce images of the entire LOFAR field-of-view using \verb|DDFacet| \citep{Tasse2018}. 

To improve the accuracy of the calibration toward the target, the two directional-dependent calibrated data sets were joined to perform an additional common calibration scheme \citep{vanWeeren2020}. This method consists of subtracting off all sources outside a small region containing the cluster, using the models derived from the pipeline, and then performing several iterations of phase and amplitude self-calibration in the extracted region to optimize the solutions at this location. In the process, we  apply the correction for the LOFAR station beam at the position of the target. 

The imaging of the LOFAR data was done in {\tt WSClean} \citep{Offringa2014}. To emphasize the radio emission on various spatial scales, we created images over a wide range of resolutions and with different uv-tapers and weighting schemes. 
In this paper, we also use VLA ( L-, S- and C-band) and uGMRT (300-850 MHz) data. For the data reduction process used for each of these auxiliary datasets, we refer to \cite{vanWeeren2016b,vanWeeren2017b,Rajpurohit2020c}. All radio observations used in the paper are summarized  in  Table\,\ref{Tabel:obs}.

The flux density uncertainties are estimated as follows:
\begin{equation}
\Delta S  = \sqrt{\left(f_{\text{scale}} \cdot S\right)^{2} + \sigma_{\text{rms}}^2 \cdot N_{\text{b}}},
\label{error}
\end{equation}
where $S$ denotes the flux density of the source, $f_{\text{scale}}$ the flux density scale uncertainty, $\sigma_{\text{rms}}$ the noise in the image and $N_{\text{b}}$ the number of beams necessary to cover the source. For the VLA data, we assume absolute flux calibration uncertainties of 4\% in L-band \citep{Perley2013} and 2.5\% in S- and C-band \citep{Perley2013}. For the LOFAR (144\,MHz) and uGMRT (Band 3 and Band 4) data, we assume a flux density scale uncertainty of 10\% \citep{Chandra2004}.

\begin{figure*}[!thbp]
    \centering
     \includegraphics[width=0.49\textwidth]{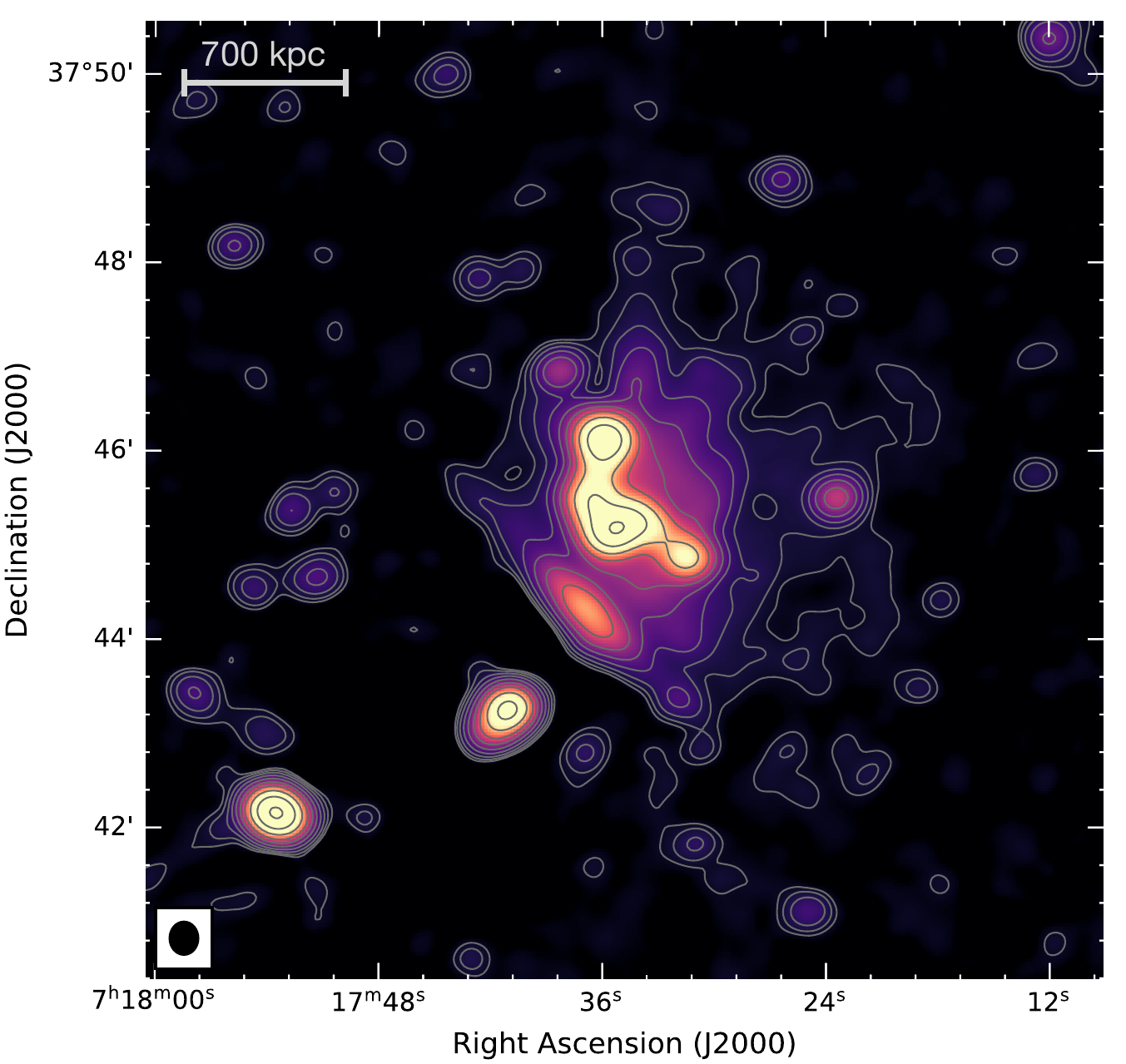}
      \includegraphics[width=0.49\textwidth]{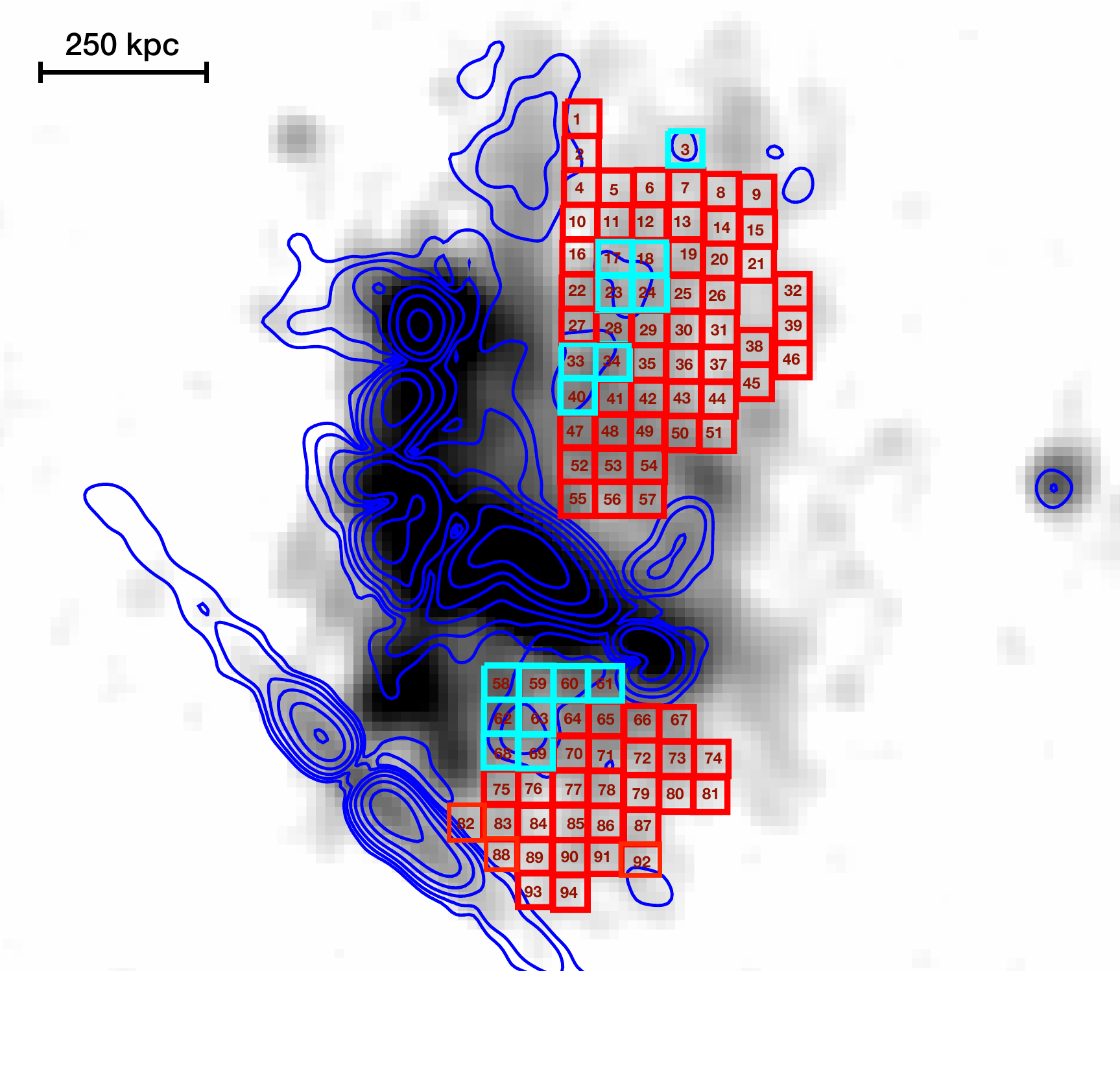}
   \vspace{-0.1cm}
    \caption{\textit{Left:} Low resolution ($20\arcsec$ ) VLA L-band image of MACS\,J0717+3745.  The low surface brightness emission visible at 144\,MHz, to the west of the cluster, is also clearly recovered in the VLA L-band image. Contour levels are drawn at $[1,2,4,8,\dots]\,\times\,3.0\,\sigma_{{\rm{ rms}}}$, where  $\sigma_{{\rm{ rms}}}=14\,\rm \upmu Jy\,beam^{-1}$. In this image, there is no region below $-4\sigma_{{\rm rms}}$. The beam size is indicated in the bottom left corner of the image. \textit{Right}: Box distribution across the halo overlaid on the LOFAR total intensity map at $8\arcsec$ resolution (gray) and polarization contours (blue) at 3\,GHz. The cyan boxes show polarized patches found in the halo region. The width of the boxes used to extract the indices is $8\arcsec$, corresponding to a physical size of about 50\,kpc.}
      \label{halo_regions}
\end{figure*}

%%%%%%%%%%%%%%%%%%%%%%%%%%%%%%%%%%%%%%%%%%%%%%%%%%%%%%%%%%%%%%%%%%%%%%%%%%%%%%%%%%%%
% Results Section
%%%%%%%%%%%%%%%%%%%%%%%%%%%%%%%%%%%%%%%%%%%%%%%%%%%%%%%%%%%%%%%%%%%%%%%%%%%%%%%%%%%%
\section{Results }
\label{results}

Fig.\,\ref{fig1} presents the LOFAR\,144\,MHz radio continuum images of MACS\,J0717.5$+$3745 created with different \textit{uv}-tapering. The new LOFAR images reveal both the previously known halo (H1+H2) and the relic emission. We label the sources as per \citet{vanWeeren2017b,Bonafede2018} (the top-left panel of Fig.\,\ref{fig1}) and append new sources to that list.

The top left panel of Fig.\,\ref{fig1} shows our deep $5\arcsec\times5\arcsec$ resolution LOFAR image, made using \verb|robust|\,$=-0.25$ weighting. The most prominent source in the image is the large chair-shaped radio relic. The image shows two distinct radio sources inside the cluster. A narrow angle tailed-galaxy (NAT) is embedded in the southern part of the relic emission. Its tails are more extended at frequencies below 700\,MHz and bends to the south of the cluster \citep{Rajpurohit2020c}. In the new LOFAR observation, we detected an additional radio galaxy, labeled as AGN. This source is only detected at 144\,MHz. 

Moving out of the cluster, there are two prominent sources, FRI and B, that are visible at all the observed frequencies; see Fig.\,\ref{compare}. The FRI is a linearly shaped foreground Fanaroff-Riley I type galaxy. Its lobes are prominent in the uGMRT and VLA images, as shown in Fig.\,\ref{compare}. The source B is a head-tail radio galaxy falling into the cluster along the large optical filament \citep{Ebeling2004}. The source A is a compact point source. 

In Fig.\,\ref{compare}, we show MACS\,J0717.5$+$3745 images from 144\,MHz to 5.5\,GHz. These images have been convolved to the common resolution of $8\arcsec$. The most striking result is that in our new LOFAR images,  the halo emission is more extended than previously seen by \cite{Bonafede2018}, in particular to the north and west of the halo (see top left and top middle panels of Fig.\,\ref{compare}). At $3.5\sigma_{\text{rms}}$, the largest linear size (LLS) of the halo is about 2.2\,Mpc at 144\,MHz.

The overall morphology of the halo emission appears similar in our LOFAR, uGMRT Band\,3/Band\,4 and VLA L-, S-, C-band images, see Fig.\,\ref{compare}. However,  the halo is more elongated to the north and northwest compared to images below 1.5\,GHz. The extended northern and the western part of the halo are not detected in the highly sensitive VLA L-, S-, and C-band images (see bottom panels of Fig.\,\ref{compare}). This may suggests that the external regions of the halo exhibit a steeper spectrum than in the central region. Similar trends have been reported for the halo in the Abell 1656 \citep[hereafter the ``Coma cluster''; ][]{Giovannini1993,Deiss1997}. 

Previously published 147\,MHz LOFAR images show that the halo emission extends toward the east of the relic \citep{Bonafede2018}, we confirm this. The eastern emission is also detected in the uGMRT  image; see top right panel Fig.\,\ref{compare}. However, we note that there are a few filamentary structures, visible in the VLA (L, S, and C band) images that emerge from the relic to the east. The eastern lobe of the FRI galaxy also seems to dominate the emission.  Therefore, the emission to the east of the relic is very likely associated with those filaments and the FRI galaxy rather than the halo.

The VLA high resolution L-, S-, and C-band images suggested the presence of several fine filamentary structures within the halo region with sizes from $100-300$\,kpc \citep{vanWeeren2017b}. Two of these filaments that extend from the relic are labeled `F1' and `F2'. The distinctions of F1 and F2 as filaments is made on the basis of the high resolution (above $3\arcsec$ resolution) VLA images as these features appear filamentary between 1 to 6.5\,GHz; see \cite{vanWeeren2017b}. At high frequencies, the boundaries of both filaments are relatively well defined. F1 also shows some hint of spectral index gradients, suggesting a shock origin \citep{vanWeeren2017b}. At $5\arcsec$ resolution, our new LOFAR image also shows significant substructure, in general, in the halo region (see top left panel of Fig.\,\ref{fig1}). In spite of the accurate characterization of the halo emission, it remains uncertain whether or not these substructures are associated with the halo or whether they are related to shocks (similar to relics). We note that in our analysis we do not consider F1 and F2 as a part of the halo as they are highly polarized between 1-6.5\,GHz (see the right panel of Fig.\,\ref{halo_regions}) and we suggest that they are not associated with emission from the halo (Rajpurohit et al. to be submitted).

\begin{figure*}[!thbp]
    \centering
    \includegraphics[width=1.0\textwidth]{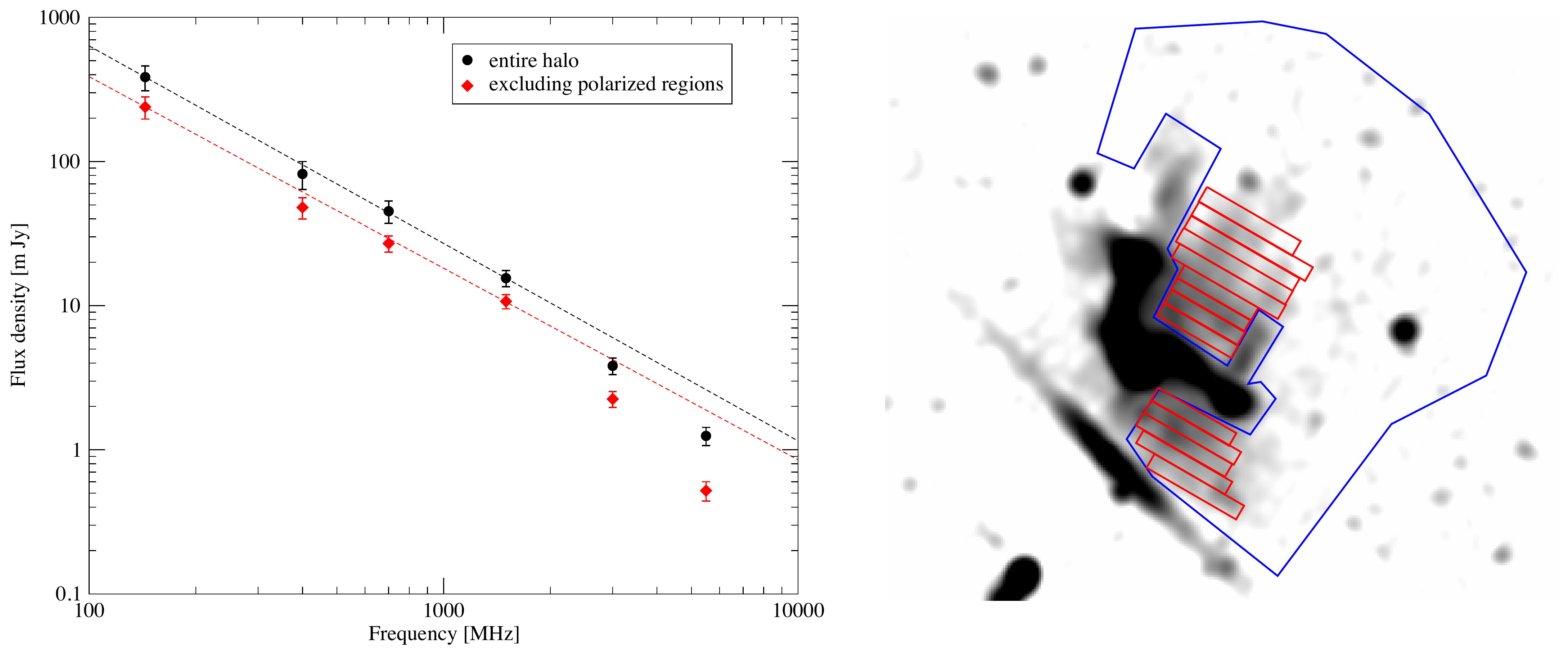}
    \vspace{-0.7cm}
    \caption{  \textit{Left:} Integrated spectrum of the halo between 144\,MHz and 5.5\,GHz. Dashed lines are fitted (between 144\,MHz and 1.5\,GHz) straight power-laws. The overall spectrum of the halo cannot be described by a single power-law spectrum. The halo spectrum is curved and shows a high frequency spectral steepening above 1.5\,GHz. Black and red points show the integrated spectrum of the halo obtained by including and excluding polarized patches, respectively, detected in the halo region. The polarized regions are shown with cyan boxes in the right panel of Fig.\,\ref{halo_regions}.  \textit{Right} VLA 1.5 GHz  image depicting the blue region where the integrated flux densities of the halo was measured (excluding source A). The flux densities are extracted from $15\arcsec$ resolution images, created using uniform weighting with a uv-cut at $0.2\rm\,k\,\uplambda$. The red boxes have a width of $8\arcsec$ used for extracting the radial spectral index profile across the halo.}
\label{halo_spectra}
\end{figure*} 

We also imaged the VLA L, S and C-band data at $20\arcsec$ resolution, using ${\tt robust}=0$. The VLA L-band low resolution is presented in the left panel of Fig.\,\ref{halo_regions}. The image reveals more  emission to the north and west at 1.5\,GHz. However, this extension of the halo emission is  not detected in the VLA S- and C-band low resolution  images. 

Another key difference with respect to the previous 147\,MHz LOFAR image is that we do not detect the $\sim1.9 \,\rm Mpc$ large arc-shaped source to the west of the cluster (identified by \citealt{Bonafede2018}). We detect only a few faint patches of faint emission coincident with the northern part of the radio-arc. We note that in the previously published 147\,MHz LOFAR images, there is  a large negative region to the east of the radio arc. \citealt{Bonafede2018} also reported a ``radio-bridge'' connecting the halo to a head-teal radio galaxy (source B) in the direction of the accreting sub-group along the intergalactic filament. In our new LOFAR images, we do not detect any significant emission at the location of the radio-bridge. 

We emphasize that our new LOFAR observations are deeper (16 hours on-source time) and more sensitive than the published LOFAR data (5 hours on-source time). The non-detection of both the radio bridge and the radio arc in our new images and the presence of a large negative region at the position of these two sources suggest that these are rather artifacts.

\setlength{\tabcolsep}{12pt}
\begin{table}
\caption{Spectral index at low and high frequency obtained from the integrated spectrum and spectral index maps.}
\centering
\begin{threeparttable} 
\begin{tabular}{ c c c c }
 \hline  \hline  
& Integrated spectral &  Mean spectral   \\ 
& index& index maps \\ 
\hline  
$\alpha_{144\,\text{MHz}}^{1.5\,\text{GHz}}$& $-1.39\pm0.04$&$-1.40\pm0.07$ \\ 
\\
$\alpha_{1.5\,\text{GHz}}^{5.5\,\text{GHz}}$ &$-1.93\pm0.04$&$-1.71\pm0.06$\\ 
\hline 
\end{tabular}
\end{threeparttable} 
\label{table1a}   
\end{table}

%%%%%%%%%%%%%%%%%%%%%%%%%%%%%%%%%%%%%%%%%%%%%%%%%%%%%%%%%%%%%%%%%%%%%%%%%%%%%%%%%%%%
\section{Spectral index and curvature analysis}
\label{discussion}
%%%%%%%%%%%%%%%%%%%%%%%%%%%%%%%%%%%%%%%%%%%%%%%%%%%%%%%%%%%%%%%%%%%%%%%%%%%%%%%%%%%%

We have used our new radio images at 144\,MHz (LOFAR), 300$-$850\,MHz (uGMRT) and 1$-$6.5\,GHz (VLA) to study the spectral characteristics of the halo emission.  All the spectral and surface brightness analysis at 144\,MHz is done with new LoTSS data only. The VLA and the uGMRT data were originally reported by \cite{vanWeeren2017b} and \cite{Rajpurohit2020c} respectively. 

The radio observations reported here were performed using three different interferometric arrays, each with their own \textit{uv}-coverage. The shortest baselines for the LOFAR, uGMRT band 3, uGMRT band 4, VLA L-band, VLA S -band, and VLA C-band data are $0.03\,\rm k\uplambda$,  $0.18\,\rm k\uplambda$,  $0.2\,\rm k\uplambda$,  $0.16\,\rm k\uplambda$,  $0.18\,\rm k\uplambda$, and  $0.2\,\rm k\uplambda$, respectively. To mitigate biases due to missing scale sizes, we created images with a common inner \textit{uv}-cut of $0.2\,\rm k\uplambda$. Here, $0.2\,\rm k\uplambda$ is the well-sampled baseline of the uGMRT and the VLA C-band data. This uv-cut is applied to the LOFAR and the VLA L- and S-band data. 

The deconvolution was performed in {\tt CASA}\footnote{\url{https://casa.nrao.edu/}}. For imaging we used multiscale clean \citep{Cornwell2008}, W-projection algorithm \citep{Cornwell2008}, and ${\tt nterms}=2$ \citep{Rau2011}. We also employed a \textit{uv}-taper to match the spectral properties of different spatial scales. The images were created using uniform weighting, which compensates best for differences in the sampling density in the \textit{uv}-plane when combining different interferometric data. We used the same \textit{uv}-cut to create spectral index and curvature maps discussed in Sect.\,\ref{index} and Sect.\,\ref{curvature}.

\begin{figure*}[!thbp]
    \centering
     \includegraphics[width=0.49\textwidth]{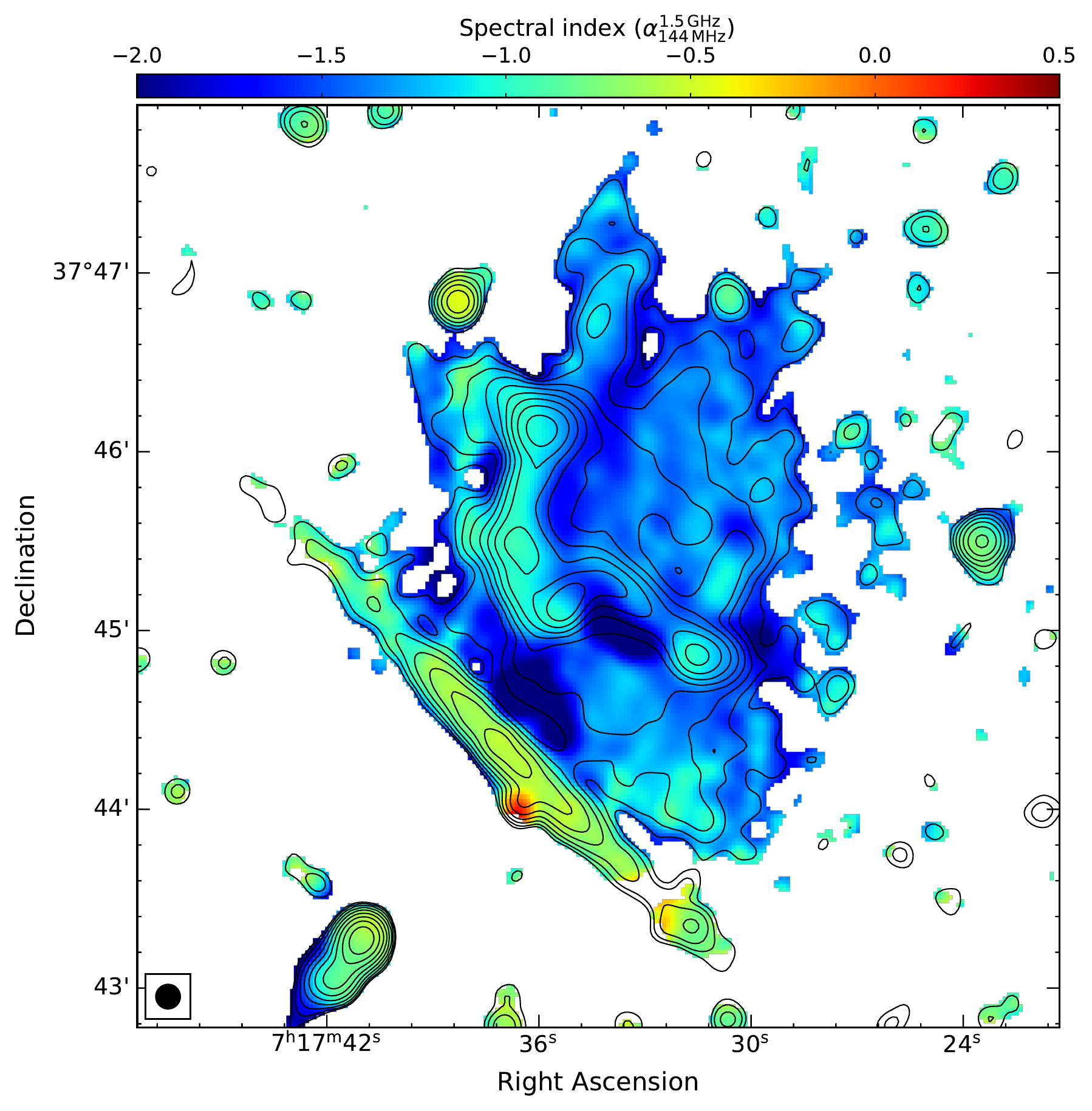}
      \includegraphics[width=0.49\textwidth]{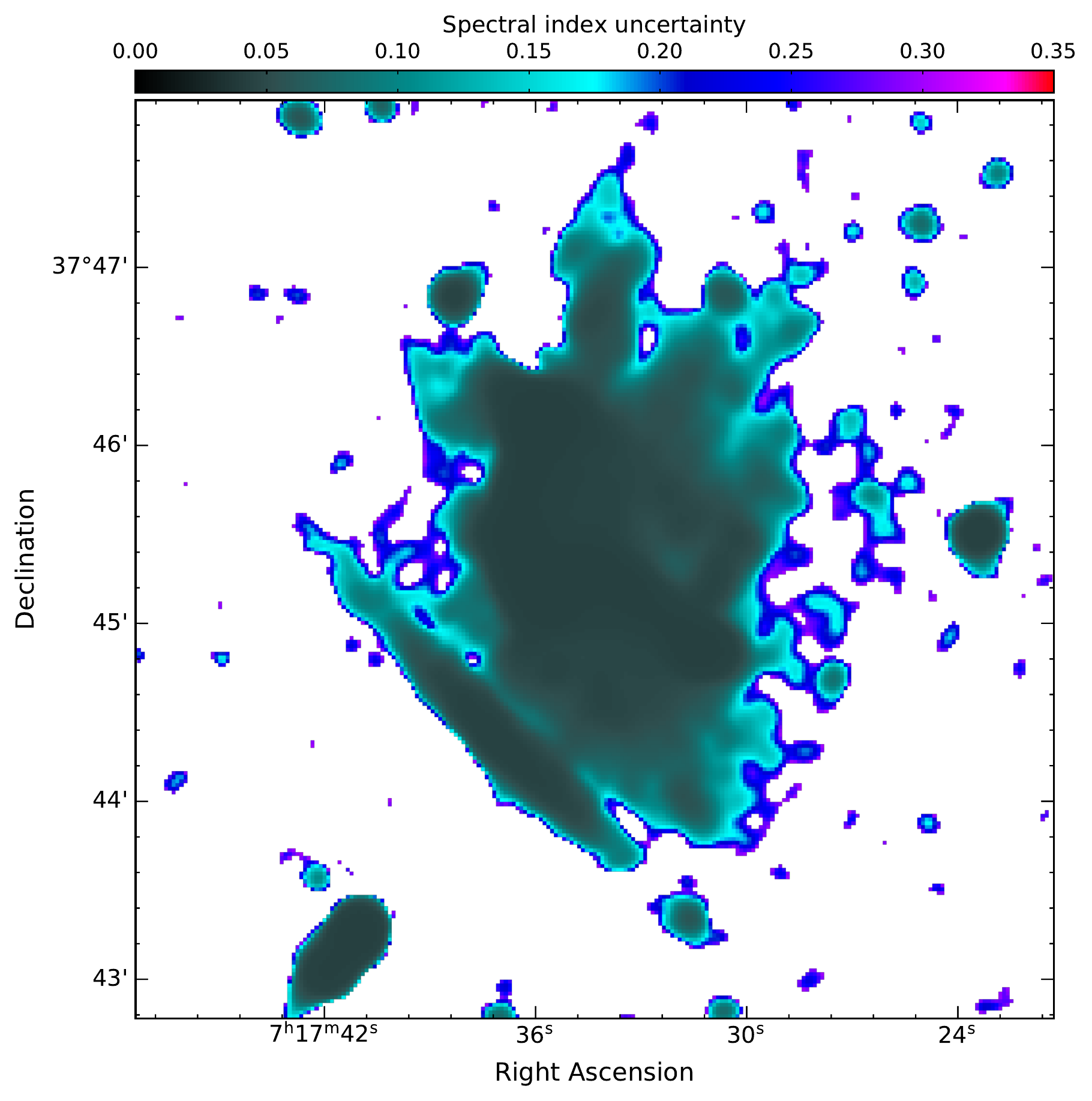}\\
      \vspace{0.2cm}
      \includegraphics[width=0.49\textwidth]{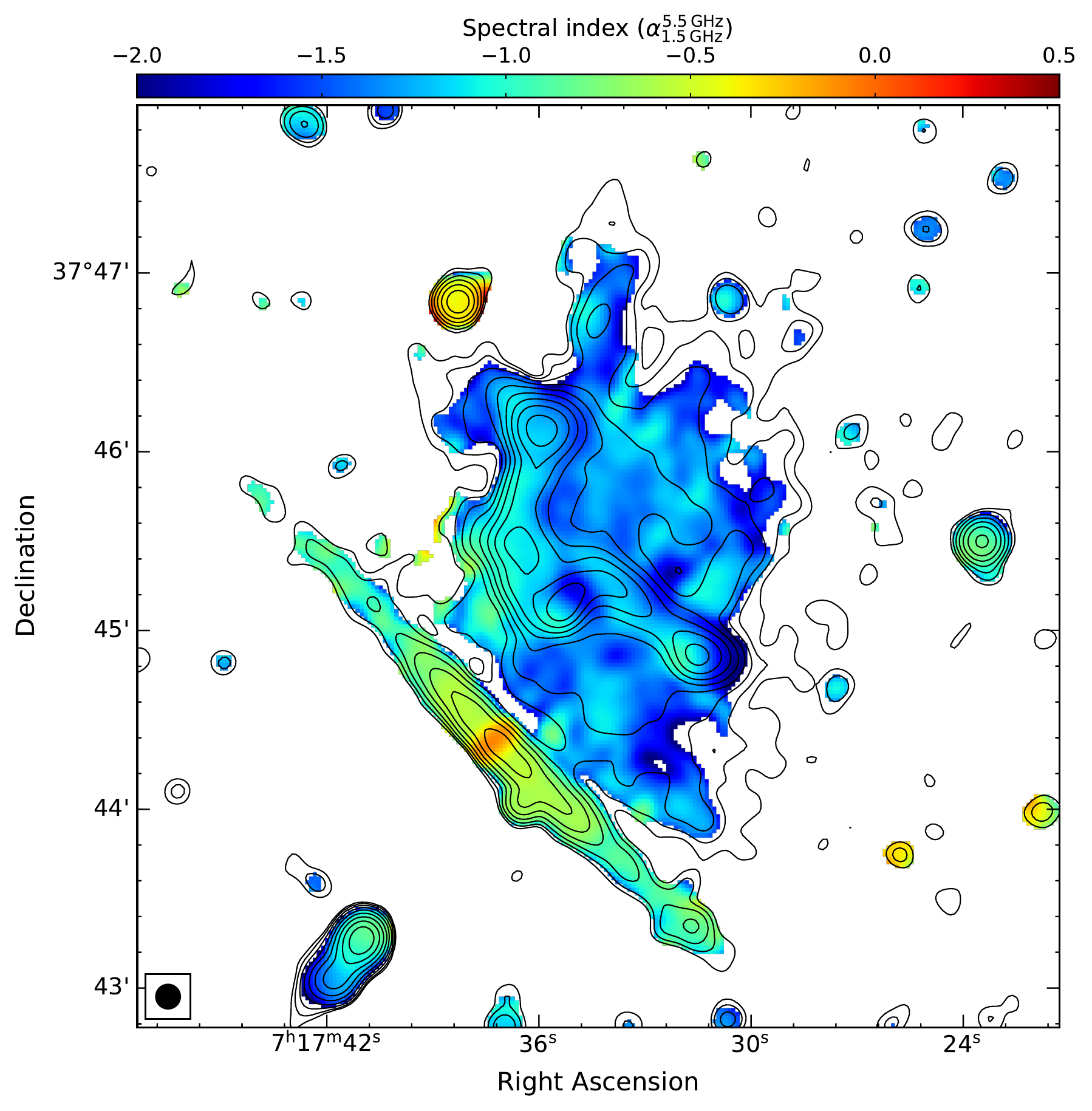}
      \includegraphics[width=0.49\textwidth]{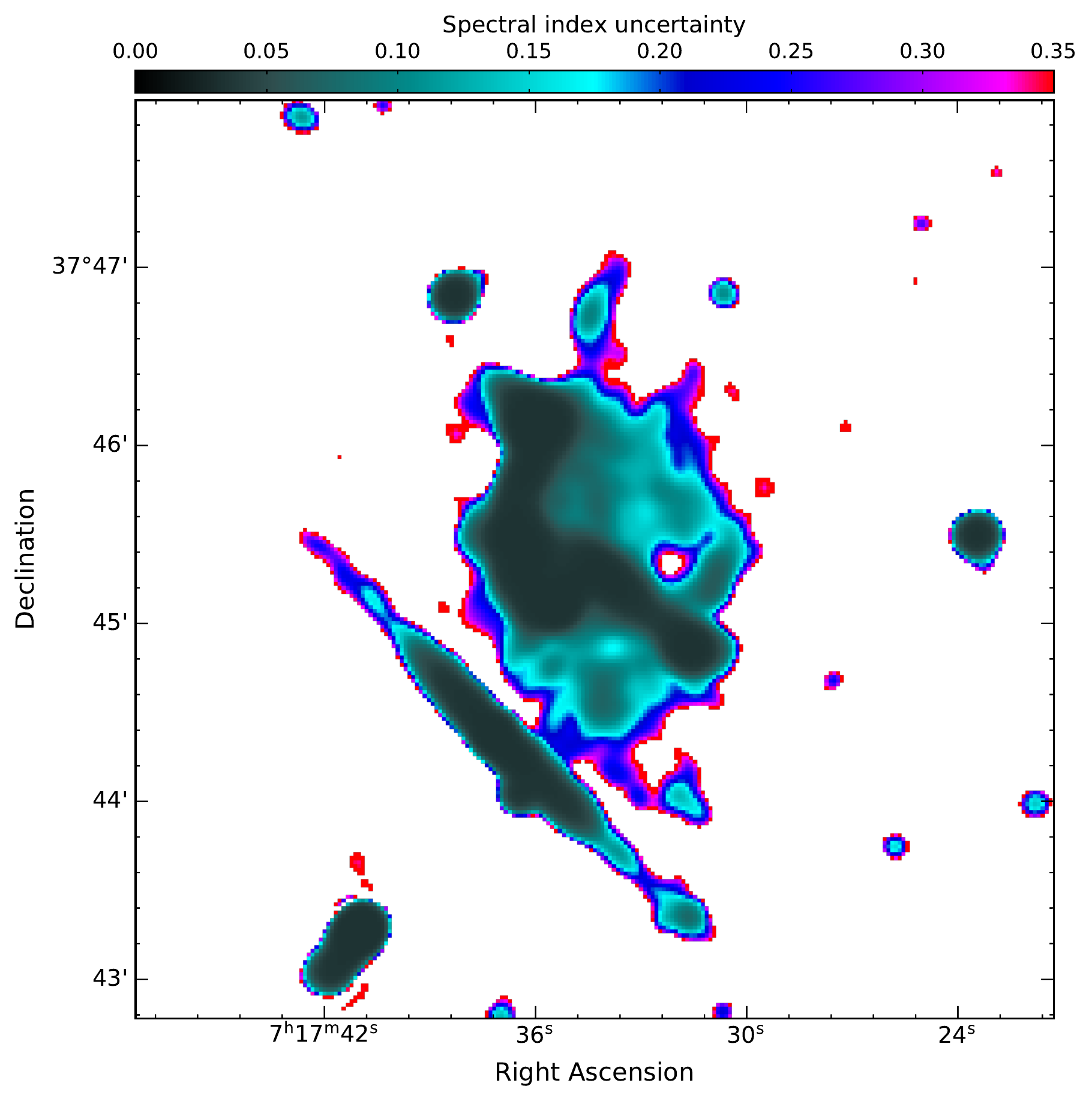}
    \vspace{-0.2cm}
    \caption{Spectral index maps of the halo in MACS\,J0717.5$+$3745 at 8\arcsec resolution. \textit{Top:} low-frequency spectral index, between 144 and 1500\,MHz, with the corresponding spectral index uncertainty. \textit{Bottom:} high-frequency spectral index between 1.5 and 5.5\,GHz, with the corresponding spectral index uncertainty. Contour levels are drawn at $[1,2,4,8,\dots]\,\times\,4\,\sigma_{{\rm{ rms}}}$ and are from the VLA L-band image. The errors are based on the individual rms noise values in the maps and an absolute flux calibration uncertainty of 2.5\%, 4\% and 10\% for the S/C-band, L-band and LOFAR (144\,MHz), respectively. The beam size is indicated in the bottom left corner of each image. It is evident that the spectral index variations across the halo are different at low and high frequencies.}
      \label{halo_index}
\end{figure*} 

\begin{figure*}[!thbp]
    \centering
     \includegraphics[width=0.49\textwidth]{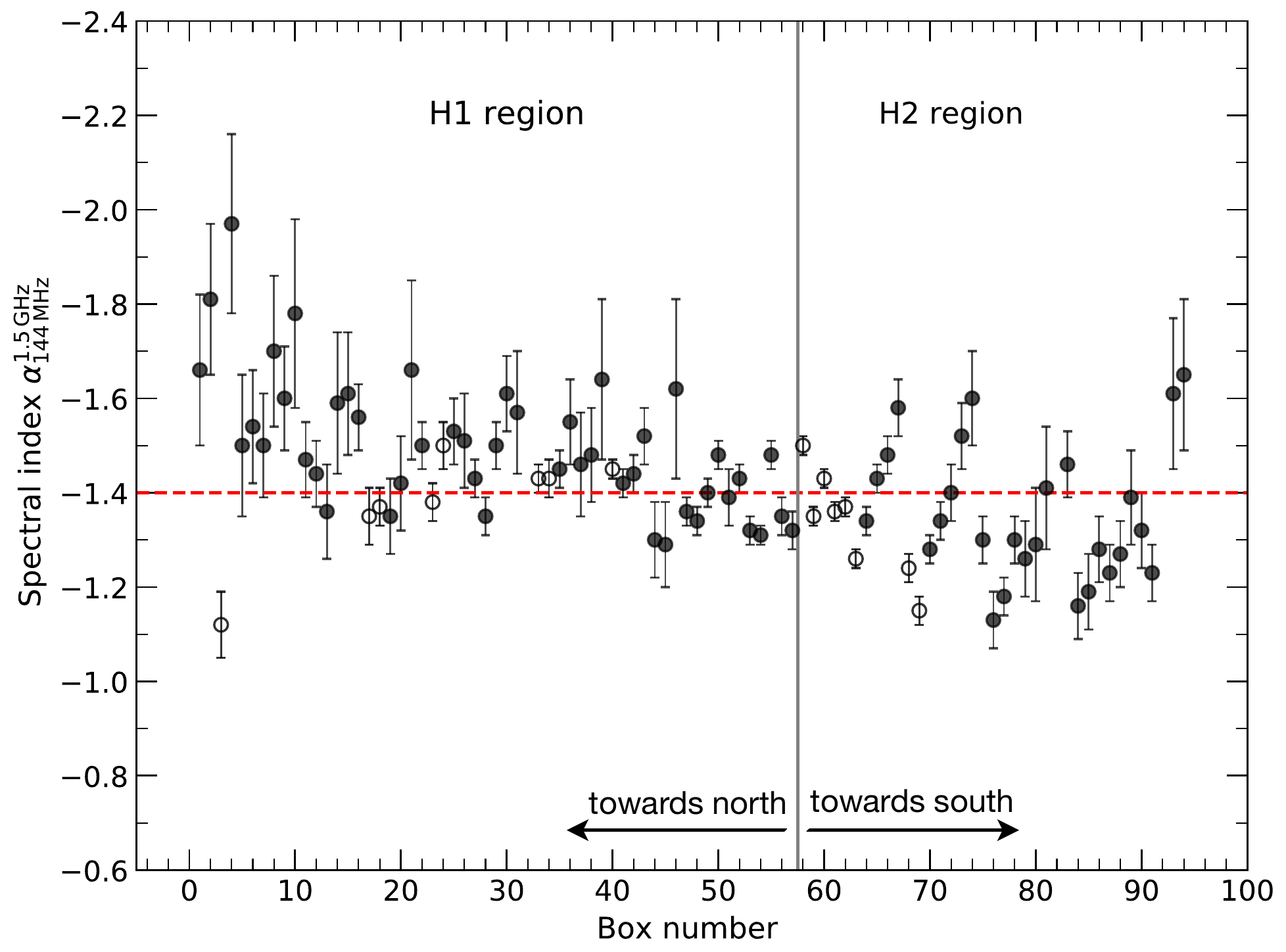}
       \includegraphics[width=0.49\textwidth]{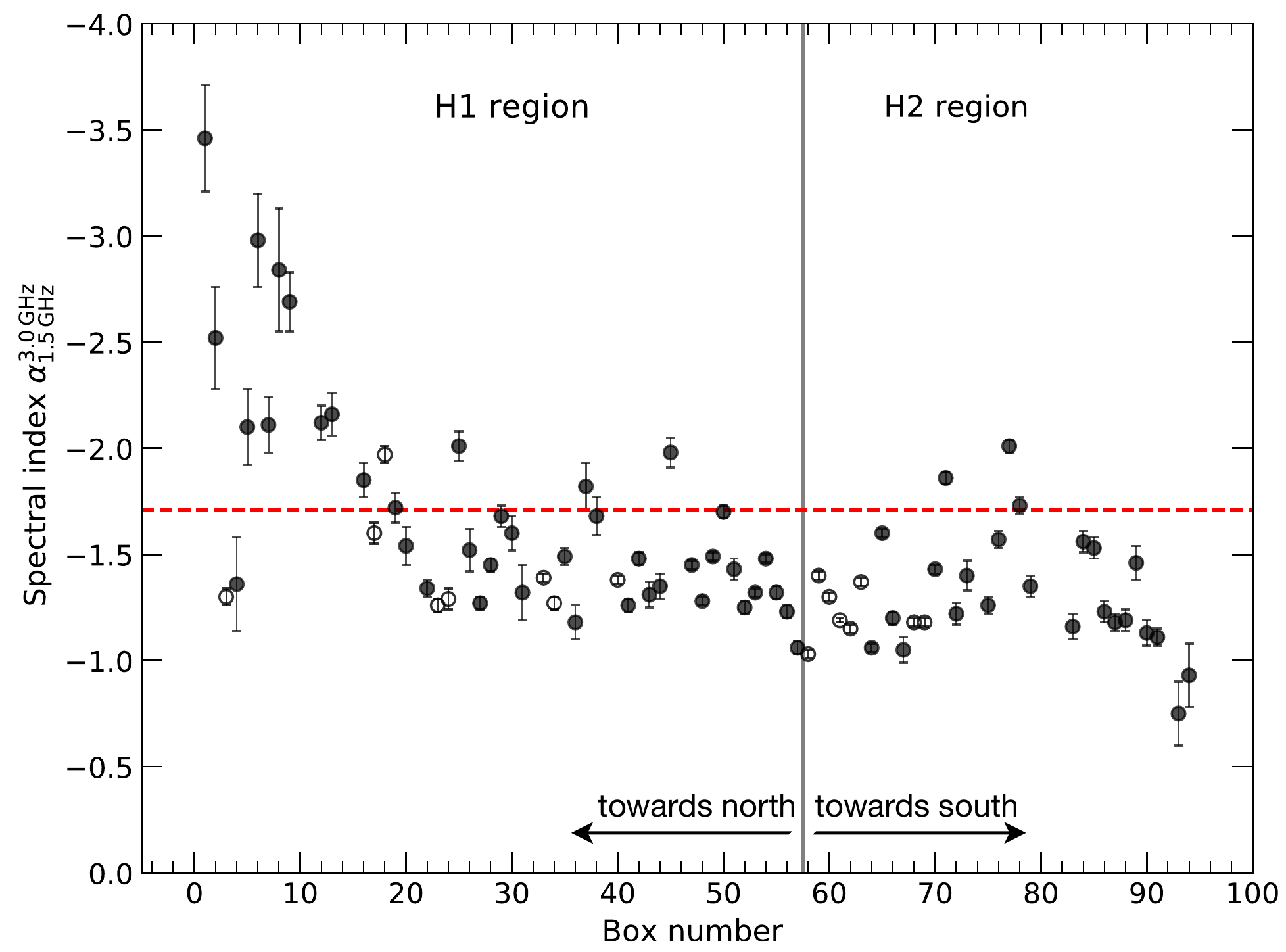}
   \vspace{-0.2cm}
    \caption{Spectral index distribution across the halo in MACS\,J0717.5$+$3745. \textit{Left}: shows the low-frequency spectral index $(\alpha_{144\,\rm{MHz}}^{1.5\,\rm{GHz}})$; \textit{Right}: shows the high-frequency spectral index $(\alpha_{1.5\,\rm{GHz}}^{3.0\,\rm{GHz}})$. The spectral indices were extracted in $8\arcsec$ boxes, corresponding to a physical size of about 50\,kpc. Open circles represent boxes within the halo where polarized emission occurs (cyan boxes in Fig.\,\ref{halo_regions}), filled circles denote unpolarized regions (red boxes in Fig.\,\ref{halo_regions}). The red horizontal line indicates the mean spectral index. The solid gray line separates the H1 and H2 region of the halo. Points close to the gray line are extracted from the central part of the halo. Systematic uncertainties in the flux-scale are included in the error bars. The spectral index tends to get steeper when moving away from the cluster center, in particular at higher frequencies and for the northern part of the halo.}
      \label{halo_dist}
\end{figure*}

%%%%%%%%%%%%%%%%%%%%%%%%%%%%%%%%%%%%%%%%%%%%%%%%%%%%%%%%%%%%%%%%%%%%%%%%%%%%%%%%%%%%
\subsection{Integrated spectrum}
\label{int}
%%%%%%%%%%%%%%%%%%%%%%%%%%%%%%%%%%%%%%%%%%%%%%%%%%%%%%%%%%%%%%%%%%%%%%%%%%%%%%%%%%%%

Radio halos show a fairly broad range of spectral indices. However, the spectral shape of halos is still poorly-known, as these studies require high-quality data and good spectral coverage, whereas many historic datasets have been limited in terms of sensitivity, dynamic range and/or \textit{uv}-coverage. MACS\,J0717.5$+$3745 is one of the few clusters that possesses a wealth of high-quality radio data across a broad frequency range, allowing us to perform a detailed spectral study. To obtain the integrated radio spectrum of the halo, we measured the flux densities from $15\arcsec$ radio maps. The region where the flux density of the halo was extracted is indicated in the right panel of Fig.\,\ref{halo_spectra}.

The resulting integrated spectrum is shown in Fig.\,\ref{halo_spectra}. There is a clear spectral break in the halo spectrum above 1.5\,GHz. The data points above 1.5\,GHz yield flux densities that are not consistent with a single power-law spectrum. The halo can be characterized by a single power-law below 1.5\,GHz, transitioning to a steeper power-law spectrum above 1.5\,GHz. A fit to the integrated  spectral index between 144\,MHz and 1.5\,GHz gives $\alpha_{\rm{low}} = -1.39\pm0.04$, a steeper value of $\alpha_{\rm{high}} = -1.93\pm0.04$ is obtained in the 1.5 to 5.5\,GHz range. This implies that the halo spectrum steepens significantly at higher frequencies. We measure the size of the halo from $15\arcsec$ resolution maps. The largest angular size of the halo at 1.5\,GHz, 3\,GHz and 5.5\,GHz is $304\arcsec$, $194\arcsec$, and $161\arcsec$, respectively. In Table\,\ref{Tabel:obs}, we listed the LAS recoverable by our radio observations. The observed sizes are much smaller than the LAS of our observations and thus they are not affected by missing spatial scales and missing short baselines. 

The spectral index of the MACS\,J0717.5$+$3745 halo is already steeper at low frequencies (144\,MHz to 1.5\,GHz) than has been measured for other well known halos (namely between $-1.05$ to $-1.25$), such as the halos in the 1RXS\,J0603.3$+$4214 \citep[aka the Toothbrush cluster;][]{Rajpurohit2019}, Coma cluster \citep{Deiss1997}, CIZA\,J2242.8$+$5301 \citep[aka the Sausage cluster;][]{Hoang2017, Gennaro2018}, and Abell S1063 \citep{Xie2020}. Since MACS\,J0717.5$+$3745 is a distant cluster, such a steep spectra might be expected due to the inverse Compton (IC) losses \citep{Cassano2010a,Cassano2010b}. We discuss this in detail in Sect.\,\ref{discussion1}.

We emphasize here that recent analysis of the wideband  VLA 1$-$6.5\,GHz data has revealed the presence of some polarized regions in the halo (Rajpurohit et al. to be submitted). These regions are denoted by the cyan boxes in the right panel of Fig.\,\ref{halo_regions}. These regions have Rotation Measure (RM) in the range $+8$ to $+20\rm \,rad\,m^{-2}$, and rather a low standard deviation of RM ($\sigma_{\rm RM}\sim 12\rm \,rad\,m^{-2}$). On the other hand, if these structures are located at the cluster center and are associated with the halo, we would expect higher RM and $\sigma_{\rm RM}$ values. Consequently, the low RM and $\sigma_{\rm RM}$ values suggest that these polarized structures embedded in the halo region are in fact related to shocks projected on the cluster center. These structures indeed affect the integrated spectrum of the halo. Hence, we also obtained the integrated spectrum of the halo by excluding the flux density contribution from those polarized regions. 

The red data points in Fig.\,\ref{halo_spectra} show the integrated spectrum of the halo when excluding the polarized filaments. The steepening towards high frequency is now even more evident. These results indicate that the MACS\,J0717.5$+$3745 halo is among the few which show a high-frequency spectral steepening. Other than the halo in MACS\,J0717.5$+$3745, there are only six radio halos where the available data cover frequencies above $\sim3.0\rm \,GHz$. These are the Coma cluster \citep{Thierbach2003}, 1E\,0657$-$56 \citep[aka the Bullet cluster;][]{Shimwell2014}, Abell 2744 \citep{Pearce2017}, the Toothbrush cluster \citep{Rajpurohit2019}, the Sausage cluster \citep{Gennaro2018}, and Abell S1063 \citep{Xie2020}. Of these seven halos, high-frequency spectral steepening has been observed in two: the halos in the Coma cluster and Abell S1063.

In principle,  the steepening of the spectrum at higher frequencies could be caused by the thermal SZ decrement. To check whether this is the case for our observation, we estimated the possible corrections for the SZ effect to the observed radio halo spectrum at high frequency  \citep[e.g.,][]{2002A&A...396L..17E,2013A&A...558A..52B}. We thus predicted the expected level of the SZ decrement for the halo in the MACS\,J0717.5$+$3745 halo, by simulating a gas density and temperature distribution using a $\beta-$model, based on \citet{2008ApJ...684..160M}, and for the reference temperature of $\approx 12.2 \rm ~keV$ \citep{vanWeeren2017b}. To compare the flux density with the same uv-coverage of our observations, we apply an inner uv-cut of $\geq 0.2 \, \rm k \uplambda$ also for the SZ calculation. We integrated the predicted SZ signal over the area corresponding to the radio halo, assuming a simple model with a uniform pressure distribution. This yielded a total SZ decrement of $-0.88$\,mJy at 5.5\,GHz, $-0.26$\,mJy at 3.0\,GHz and $-0.06$\,mJy at 1.5\,GHz. 

We emphasize here that the halo is elongated and thus has a flat brightness profile, therefore the most of the halo flux density comes from the regions where the pressure (SZ) is small. Since the halo also seems to be seen in projection with the relic emission, therefore we simply exclude the region of the bright relic when measuring the radio flux density of the halo at all observed  frequencies. In contrast, the SZ flux density estimates include those location, thus making the ratio between SZ and radio flux biased high.  While our simple model provides only a rough estimate of the SZ decrement, these results suggest that the contamination is at most $\sim65\%$ of the observed halo flux at 5.5\,GHz. The 65\% SZ correction at 5.5 GHz is still smaller than would be required to make the 5.5\,GHz flux consistent with the spectrum at low frequencies. Most importantly, the steepening is clear already in the frequency range 1.5-3\,GHz where the SZ contribution is almost negligible. This allow us to conclude that the SZ effect cannot fully account for the observed high-frequency spectral steepening.

The halo in MACS\,J0717.5$+$3745 is reported to be the most powerful halo \citep{vanWeeren2009}. We measure the flux density of $S_{\rm halo, \rm 1.4} = 16 \pm 2.0 \,\rm mJy$ (including polarized regions). The monochromatic radio power of the halo is derived as: 
\begin{equation}
P_{1.4\,\rm GHz} =4\pi D^{2}_{L}S_{\nu_{0}} (1+z)^{-(1+\alpha)},  
\end{equation}
where $D_{L}$ is the luminosity distance to the source, $\alpha$ is the spectral index used in the $k$-correction, $S_{\nu_{0}}$ is the flux density at $\nu_{0}=1.4\,\rm GHz$.

The L-band flux density corresponds to a rest-frame luminosity of $P_{1.4\rm \, GHz} = (2.2\pm0.3) \times 10^{25} \,\rm W \, Hz^{-1}$ (adopting the spectral index of $\alpha=-1.39$). Our estimated radio power of MACS\,J0717.5$+$3745 is in agreement with the radio power-vs.-mass and radio power-vs.-X-ray luminosity scaling correlations of the known radio halos \citep[e.g.,][]{Cassano2013}.

We also obtained the radio power by excluding the polarized regions seen in the halo, namely $P_{1.4\,\rm GHz, \,lower\,limit} = (1.4\pm0.2) \times 10^{25} \,\rm W \, Hz^{-1}$. Since the halo is very likely seen in projection with the relic, this is the lower limit. To estimate the upper limit of the halo, we  extrapolated the halo flux at the location of the relic using the average flux from the halo per unit surface area outside the relic emission. This gives the radio power  of $P^{}_{1.4\,\rm GHz, \,upper\,limit} = (3.0\pm0.3) \times 10^{25} \,\rm W \, Hz^{-1}$. This value is a bit lower compared to \cite{vanWeeren2009} due to the fact that they used less sensitive GMRT and VLA observations. In addition, they obtained the radio power at 1.4\,GHz by extrapolating the narrow-band GMRT 610\,MHz data and adopting a spectral index of $-1.10$ in an area of 1.2\,Mpc.

%%%%%%%%%%%%%%%%%%%%%%%%%%%%%%%%%%%%%%%%%%%%%%%%%%%%%%%%%%%%%%%%%%%%%%%%%%%%%%%%%%%%
\subsection{Spectral index maps}
\label{index}
%%%%%%%%%%%%%%%%%%%%%%%%%%%%%%%%%%%%%%%%%%%%%%%%%%%%%%%%%%%%%%%%%%%%%%%%%%%%%%%%%%%%

Spatially-resolved spectral index mapping of halos can provide useful information on their origin and connection with the merger processes. Our VLA and LOFAR images allow us to derive a spectral index map for the observed radio halo at high resolution and with high accuracy.

In \cite{vanWeeren2017b}, high resolution spectral index maps (at $5\arcsec$ and $10\arcsec$ resolutions) of the halo were derived using the VLA 1.5$-$5.5\,GHz data. Recently, \cite{Bonafede2018} presented low frequency (147$-$610\,MHz) spectral index maps of the halo (at $10\arcsec$ and $30\arcsec$ resolutions).  The new LOFAR data allow us to reconstruct the surface brightness distribution at 144\,MHz with adequate resolution and improved sensitivity.

We created LOFAR and VLA images of the halo at $8\arcsec$ resolution, using the same imaging parameters as described in Sect.\,\ref{int}. This resolution was chosen as a compromise between providing the resolution necessary to probe different substructures, while retaining high signal-to-noise (SNR). From our high-resolution data, we know that there are only a few extended or point-like discrete sources embedded in the halo region; as their contamination is minimal, they were not subtracted. However, these sources can be clearly visually separated from the halo emission. After imaging, we ensured both datasets were on the same grid (using the \verb|CASA| task {\tt imregrid}) before masking each image at the $4\sigma_{{\rm{ rms}}}$ level. Finally, we derived pixel-wise spectral index maps at both low frequency (144\,MHz to 1.5\,GHz) and high frequency (1.5 to 5.5\,GHz).

To obtain the spectral index uncertainty map, we take into account the image noise and the absolute flux calibration uncertainty. We estimate the spectral index error via:
\begin{equation}
\Delta{\alpha}  =  {\frac{1}{\ln\left( {\frac{\nu_1}{\nu_2}}\right)}} {\sqrt{{{\left({\frac{\Delta S_1}{S_1}}\right)}^2 + { { \left({\frac{\Delta S_2}{S_2}}\right)}^2}}}}
\end{equation}
where $S_1$ and $S_2$ are the flux density values at each pixel at the respective frequencies.

Fig.\,\ref{halo_index} presents the low-frequency (top) and high-frequency (bottom) spectral index maps, along with the corresponding uncertainty. We note that the spectral index uncertainties across the halo are small, mainly in the range (3-15)\%. At $8\arcsec$ resolution, these are among one of the best resolved spectral index maps of a radio halo. There are significant local variations in the spectral index across the halo. Moreover, the spectral index variations across the halo region appear different at low and high frequency.

\begin{figure}[!thbp]
    \centering
     \includegraphics[width=0.5\textwidth]{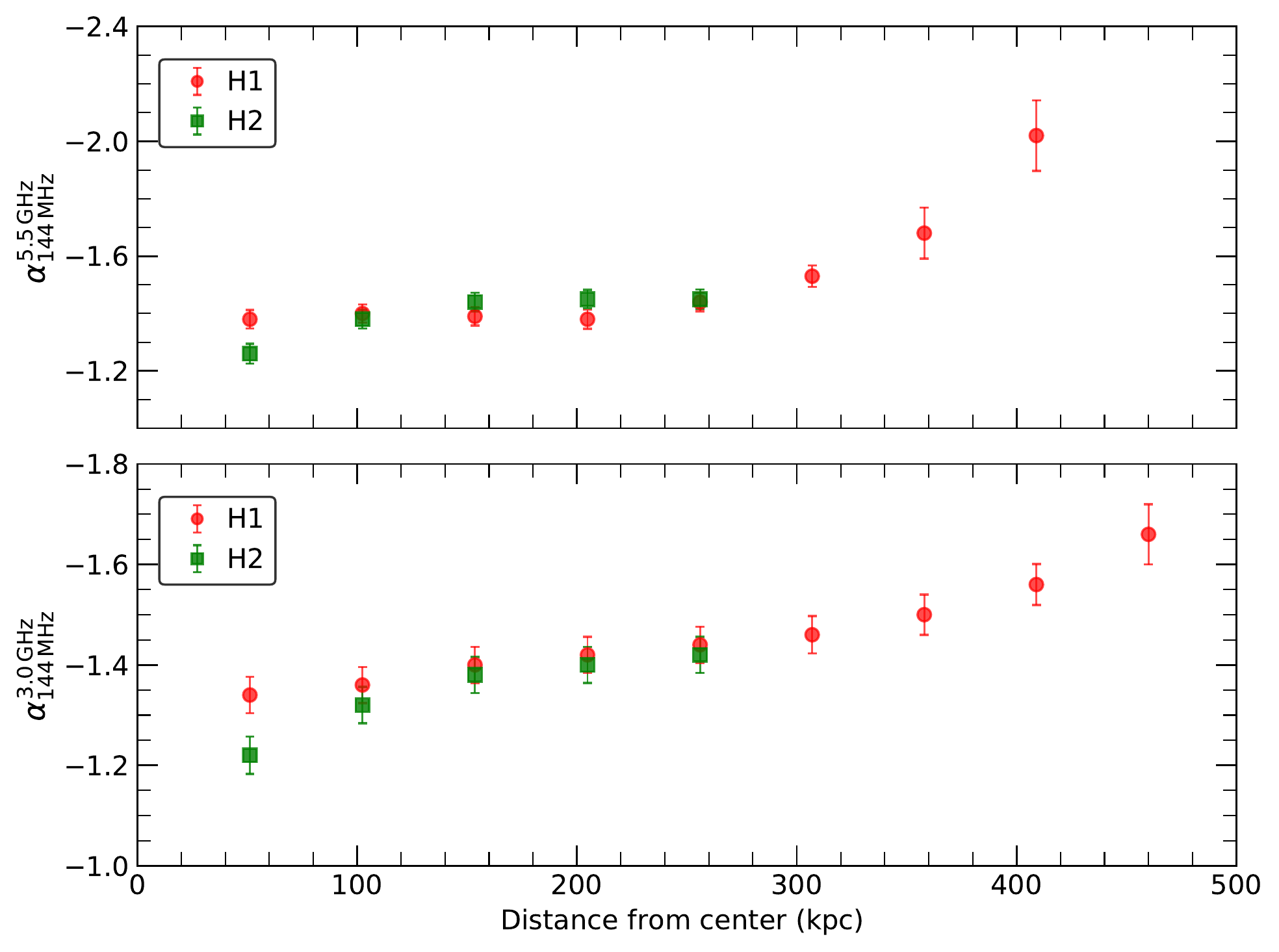}
     \vspace{-0.7cm}
    \caption{ Spectral index profiles across the radio halo from 144\,MHz and 5.5\,GHz. The spectral indices were extracted in boxes with a width of $8\arcsec$, corresponding to a physical size of about 50\,kpc. There is indeed a clear radial steepening in the spectral index value of the radio halo for both the H1 and H2 region of the halo.}
      \label{halo_profile}
\end{figure}

\begin{figure*}[!thbp]
    \centering
     \includegraphics[width=0.49\textwidth]{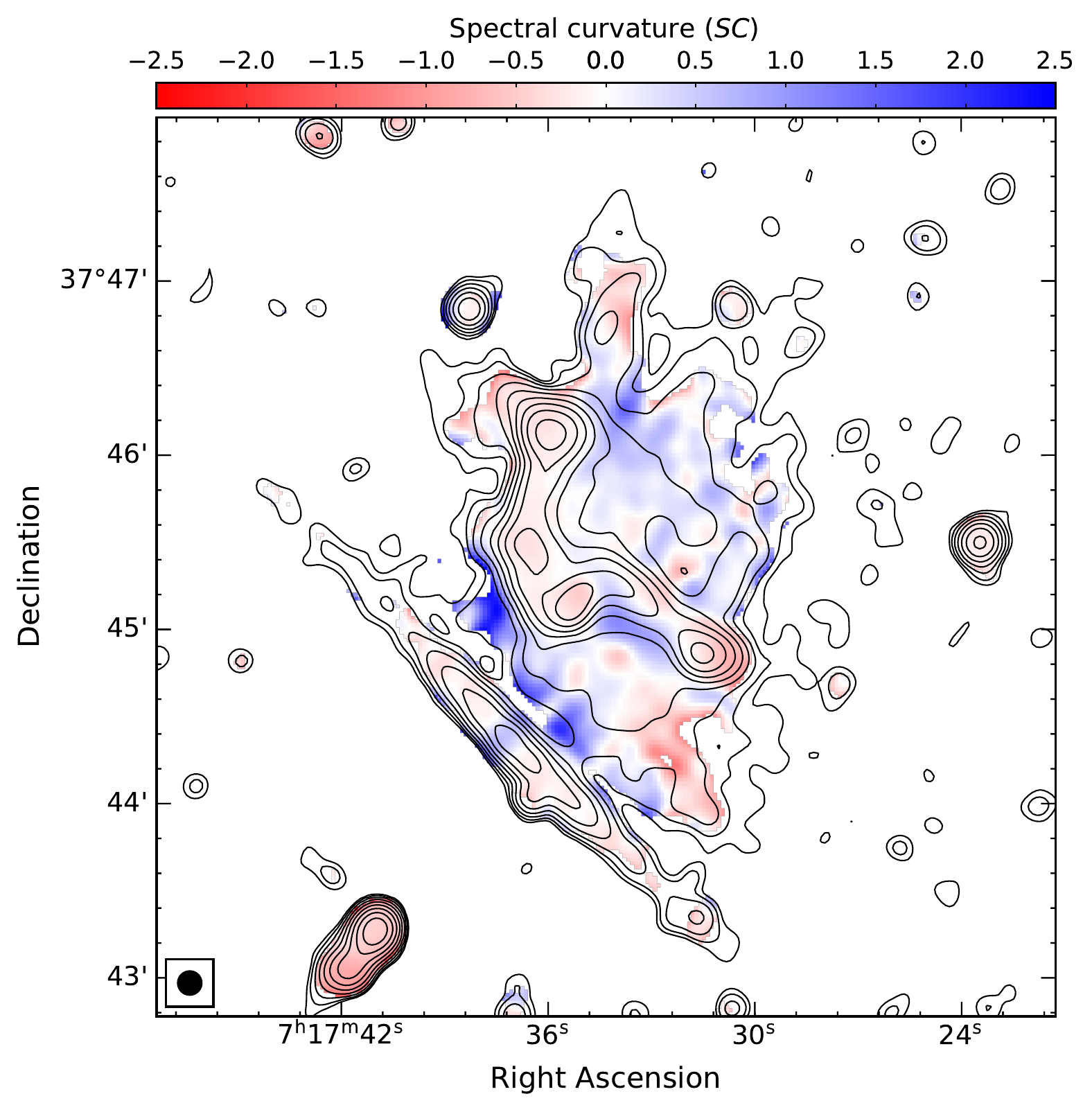}
      \includegraphics[width=0.49\textwidth]{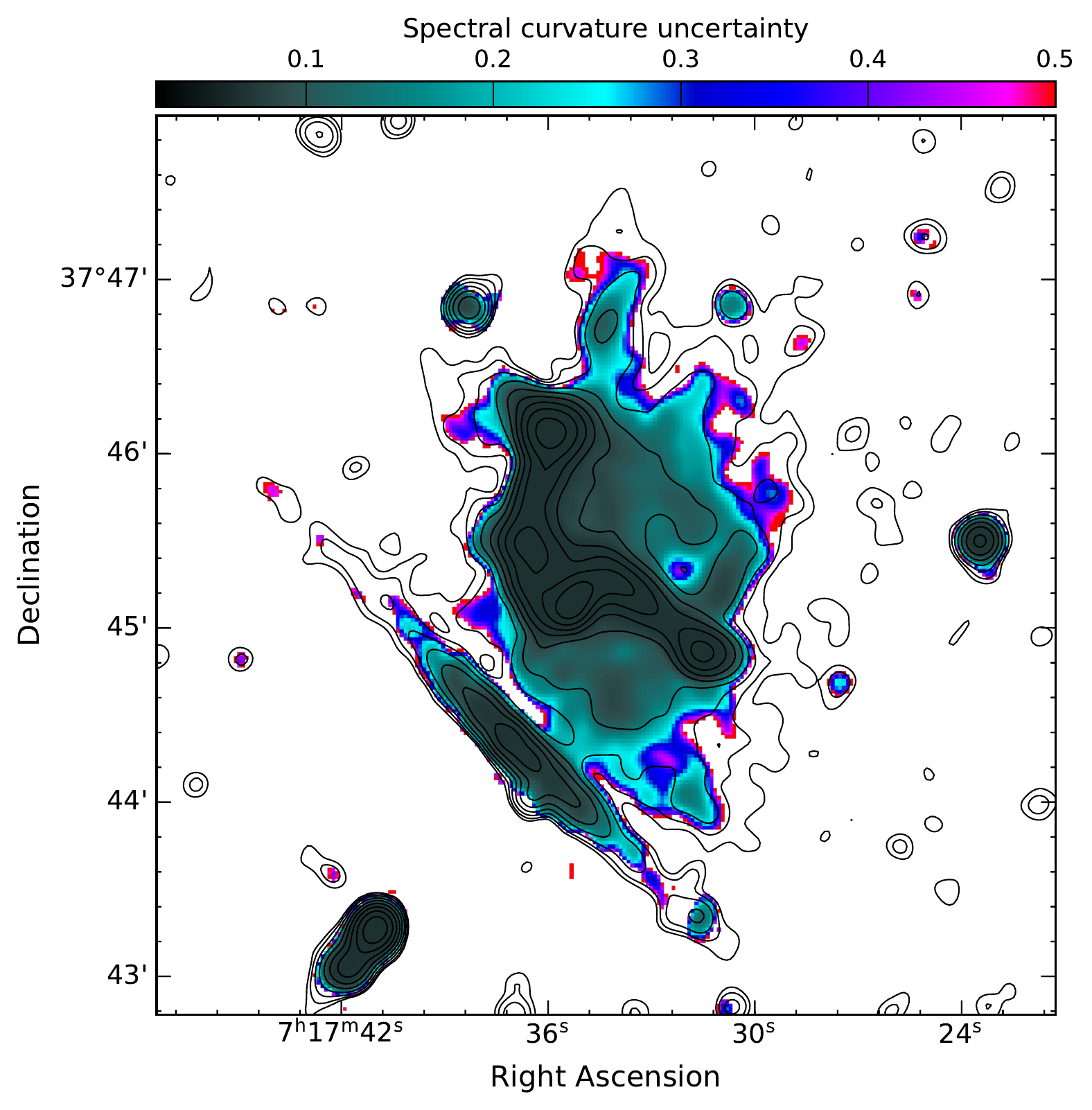}
    \vspace{-0.2cm}
    \caption{Spectral curvature map of the halo in MACS\,J0717.5$+$3745 at $8\arcsec$ resolution. \textit{Left:} Four frequency spectral curvature map between 144\,MHz and 5.5\,GHz. The SC is negative for a convex spectrum. From the curvature map it is evident that the halo shows a clear curvature. There are also significant local variations in the curvature. \textit{Right:} Corresponding uncertainty between 1.5 and 5.5\,GHz.  Contour levels are drawn at $[1,2,4,8,\dots]\,\times\,4\,\sigma_{{\rm{ rms}}}$,  and are from the VLA L-band image.}
    \label{SC}
\end{figure*}

To investigate the spatial distribution of the spectral index across the radio halo, we extracted the spectral index in several regions as indicated in the right panel of Fig.\,\ref{halo_regions}. These regions correspond to boxes with a physical sizes of about 50\,kpc. Cells with emission above $3\sigma$ in the radio images are retained for the analysis. The resulting plots are displayed in Fig.\,\ref{halo_dist}. 

The most striking result is that the spectral index distribution in the central regions (box numbers 40 to 70 in Fig.\,\ref{halo_dist}) is relatively constant ($\alpha \sim -1.35$). The spectral index steepens toward the outermost regions of the halo, reaching values $\alpha \lesssim -2.0$. The steepening in outer regions of the halo is most evident at high frequency; see the right panel of Fig.\,\ref{halo_dist}. For instance, in the northern part of the halo, the low-frequency spectral index is typically $-2.0 \lesssim \alpha_{144\,\rm{MHz}}^{1.5\,\rm{GHz}} \lesssim -1.3$, whereas the high-frequency spectral index is typically $-3.5 \lesssim \alpha_{1.5\,\rm{GHz}}^{3.0\,\rm{GHz}} \lesssim -1.2$.  Spectral steepening in the outermost regions has also observed in the halo in Coma cluster \citep{Giovannini1993,Deiss1997}.  

From our spatially-resolved spectral index map between 144\,MHz and 1.5\,GHz, we measure a mean halo spectral index is about $-1.4$ (see left panel of Fig.\,\ref{halo_dist}). At higher frequencies, between 1.5\,GHz and 3\,GHz, the mean spectral index is about $-1.7$ (see right panel of Fig.\,\ref{halo_dist}). We note that this only applies to regions of the halo that are detected clearly at high SNR from 144\,MHz to 3\,GHz. The halo extends further to the north and northwest below 1.5\,GHz, however these regions are not visible above 1.5\,GHz, thus these regions are excluded from the right panel of Fig.\,\ref{halo_dist}. However, we can conclude that the spectral index in these regions is likely steeper than $-1.7$. 

We also derived a radial profile of the average spectral index value extracted from the red rectangular boxes shown in the right panel of Fig\,\ref{halo_spectra}. The resulting profiles are shown in Fig\,\ref{halo_profile}. There is indeed a clear trend of radial steepening in the spectral index value of the radio halo, in particular between 144\,MHz and 5.5\,GHz . 

In Table\,\ref{table1a}, we summarize results obtained for the spectral index estimates from the integrated spectrum and the spatially resolved spectral index maps at low and high frequency. The mean low-frequency (144\,MHz to
1.5\,GHz) spectral index ($-1.40$) is consistent with that measured from our integrated spectrum ($-1.39\pm0.04$). However, the high frequency (1.5 to 5.5\,GHz) mean spectral index (about $-1.7$) is different than the integrated spectral index ($ -1.93\pm0.04$). As mentioned above, this is due to the fact that the halo emission is more extended at 1.5\,GHz compared to 5.5\,GHz, and consequently some regions are not included in the spatially resolved spectral index maps.

To estimate the measurement uncertainties, we followed \cite{vanWeeren2016a}. We find a mean scatter of 0.28 and 0.53 around the mean spectral index between 144$-$1500\,MHz and 1.5$-$3\,GHz, respectively. This scatter is larger compared to other well sampled radio halos (the Toothbrush cluster, Abell 2744, and Abell 520). For the halo in the Toothbrush cluster, the intrinsic scatter is remarkably low \citep[$0.04$;][]{vanWeeren2016a}. For the radio halo in A2255, \cite{Botteon2020} reported a scatter of $\sim0.3$ between 144\,MHz to 1.2\,GHz. The scatter in MACS\,J0717.5$+$3745 halo is even larger than that of the halo of Abell 2255. We also checked the spectral index distribution by adopting a larger cell size. With a cell size of about $15\arcsec$ (90\,kpc), we find that the scatter decreases slightly, for example between 1.5\,GHz to 5.5\,GHz the mean scatter decreases from 0.53 to 0.48 with a cell size of 50 and 90\,kpc, respectively.  Similar trends are reported recently for the Abell 2255 halo by \citet{Botteon2020}.

\begin{figure*}[!thbp]
    \centering
       \includegraphics[width=0.49\textwidth]{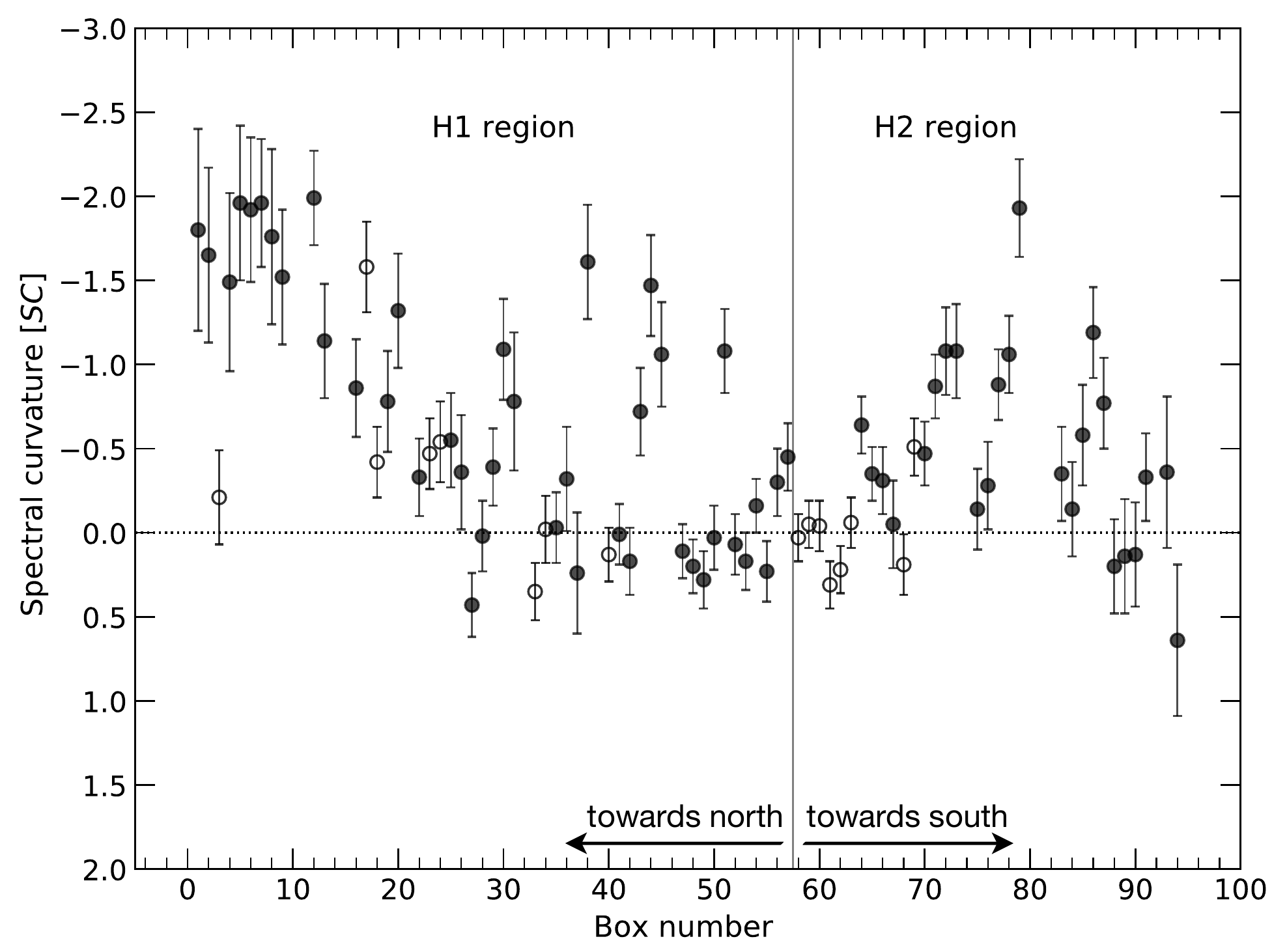}
         \includegraphics[width=0.45\textwidth]{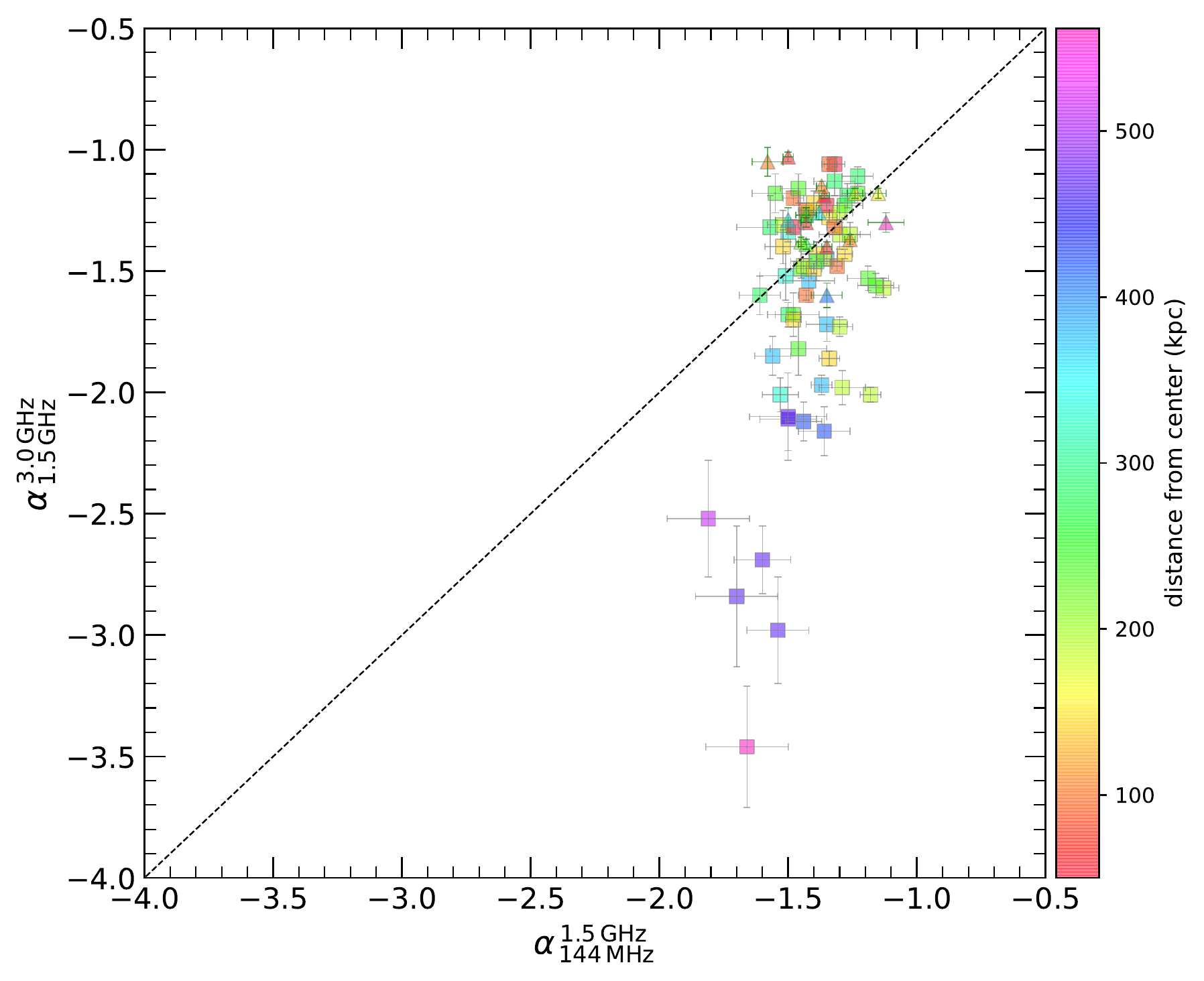}
         \vspace{-0.2cm}
           \caption{Curvature distribution and the radio color-color plot of the halo in MACS\,J0717.5$+$3745. \textit{Left}: Spectral curvature distribution across the radio halo from 144\,MHz and 3.0\,GHz. Cells with emission above $3\sigma$ in the radio images are considered. The open circles represent polarized emission in the halo region. The horizontal line indicates zero curvature. The solid gray line separates the H1 and H2 region of the halo. Points close to the gray line are extracted from the central part of the halo. The halo shows a clear trend of increasing curvature from center to the outermost regions. \textit{Right}: Radio color-color plot of the halo. The halo shows a clear negative curvature. The triangle represents polarized regions embedded  in the halo region. Some of the data points lies significantly above the power law line, suggesting the possible superposition of the halo and the relic emission.  The color bar shows the distance from the cluster center in kiloparsecs.}
      \label{halocc_plot}
\end{figure*}

%%%%%%%%%%%%%%%%%%%%%%%%%%%%%%%%%%%%%%%%%%%%%%%%%%%%%%%%%%%%%%%%%%%%%%%%%%%%%%%%%%%%
\subsection{Spectral curvature maps}
\label{curvature}
%%%%%%%%%%%%%%%%%%%%%%%%%%%%%%%%%%%%%%%%%%%%%%%%%%%%%%%%%%%%%%%%%%%%%%%%%%%%%%%%%%%%

Using our multifrequency data, we also derive a spectral curvature map to investigate whether the halo shows any curvature. The spectral curvature (SC) can be obtained using data at four frequencies, and is derived as:
\begin{equation}
{\rm{SC}} = -\alpha_{\nu_2}^{\nu_1}+\alpha_{\nu_4}^{\nu_3},
\end{equation}
where $\nu_1$ (144 MHz) is the lowest frequency and $\nu_3$ (5.5\,GHz) is the highest one. A value of ${\rm{SC}}=0$ indicates no curvature. By this convention, curvature is negative for a convex spectrum, and positive for a concave spectrum. We created the curvature map using 144\,MHz, 1.5\,GHz, 3\,GHz, and 5.5 GHz. The low frequency spectral index map is created between map 144\,MHz and 1.5\,GHz  while the high frequency between 3.0\,GHz and 5.5\,GHz. All pixels below the $4\sigma_{\rm rms}$ level were blanked. The corresponding uncertainty in SC is defined as:
\begin{equation}
\Delta\,{\rm{SC}}=\sqrt{(\Delta \alpha_{\rm low})^2+(\Delta\alpha_{\rm high})^2}.
\end{equation}
We present our curvature map in Fig.\,\ref{SC}. Both the H1 and H2 regions exhibit spectral curvature, with local fluctuations in the curvature. In general, the curvature is mainly negative (convex) and reaches about $-2$ across the halo. However, there are also regions with positive curvature (concave).

We perform the spatial distribution of SC using the regions shown in the right panel of Fig.\,\ref{halo_regions}. The left panel of Fig.\,\ref{halocc_plot} shows the halo spectral curvature distribution from 144\,MHz to 3.0\,GHz. We do not use the C -band data for studying the curvature distribution because the total extent of the halo decreases at high frequencies and this would significantly reduce the area over which we can study the curvature. As with the halo spectral index distribution, in the innermost regions  (close to the solid gray line), the curvature distribution remains largely constant. In the outermost regions, the spectrum becomes increasingly curved, reaching  $-2.0$.

To investigate this further, we make use of a color-color diagram \citep[as used in, for example,][]{Rajpurohit2019}. The halo color-color plot is shown in the right panel of Fig.\,\ref{halocc_plot}. This provides similar information to the curvature map; any point that lies away from unity (the dashed line, which denotes power-law behavior) indicate curvature. From Fig.\,\ref{halocc_plot}, we find clear evidence of spectral curvature already between 144\,MHz and 3.0\, GHz: while some regions lie close to the power-law line (indicating little-to-no curvature), the majority of regions within the halo lie away from unity. These regions tend to lie below the power-law line, indicating a clear negative curvature.

Examining the spatial distribution of these regions shows that the points that cluster around the power-law line tend to be located in the innermost regions of the halo, while the increasing curvature is seen in the outermost regions of the halo. To our knowledge, such a well-(spatially-)resolved spectral curvature study is only available for one other radio halo: the Toothbrush cluster \citep{Rajpurohit2019}. The curvature distribution  across the halo in MACS\,J0717.5$+$3745 is different than that of the halo in the Toothbrush cluster, which rather shows a uniform spectral index distribution and no sign of curvature \citep{Rajpurohit2019}.

The right panel of Fig.\,\ref{halocc_plot} also shows data points that lie significantly above the power-law line, indicating a concave spectrum. This is expected when radio emission from two different regions overlap. These regions are mostly polarized and very likely related to shock waves projected on the halo emission. The concave points in the color-color plot suggest that both halo and relic emission is present, but in different proportions at different frequencies.

%%%%%%%%%%%%%%%%%%%%%%%%%%%%%%%%%%%%%%%%%%%%%%%%%%%%%%%%%%%%%%%%%%%%%%%%%%%%%%%%%%%%
\section{X-ray and radio comparison }
%%%%%%%%%%%%%%%%%%%%%%%%%%%%%%%%%%%%%%%%%%%%%%%%%%%%%%%%%%%%%%%%%%%%%%%%%%%%%%%%%%%%

All formation models of radio halos predict a connection between X-ray and radio emission \citep[e.g.,][]{Brunetti2014}. Radio emission from halos typically follows the X-ray emission from the thermal gas \citep[e.g.,][]{Govoni2001b}. However, there are also a few clusters in which the halo emission does not clearly trace the X-ray emission \citep{Giacintucci2005}. The close morphological similarity between radio halo emission and X-ray emission suggests an interplay between thermal and non- thermal components in the ICM.

\begin{figure*}[!thbp]
    \centering
    \includegraphics[width=0.49\textwidth]{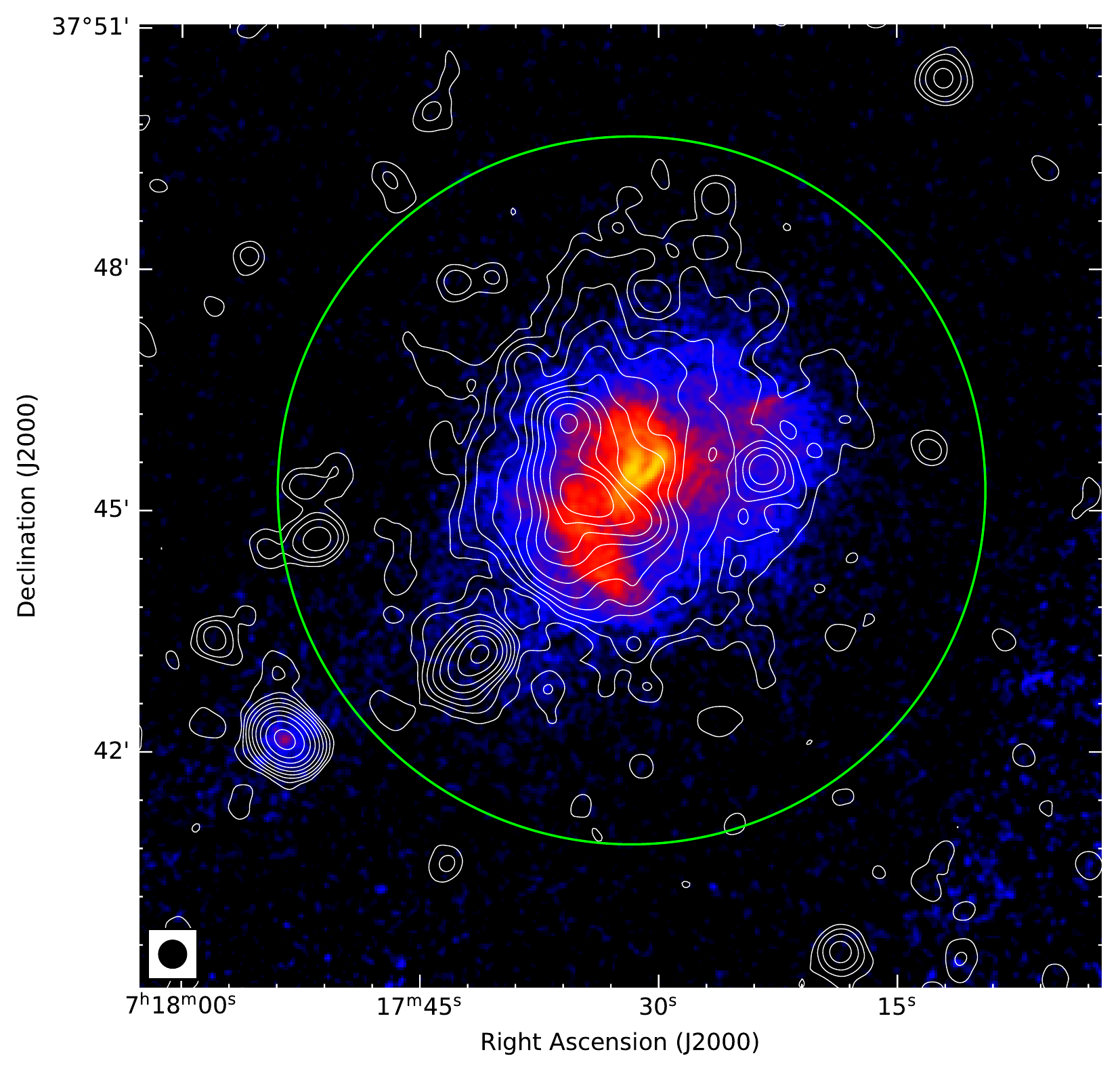}
      \includegraphics[width=0.49\textwidth]{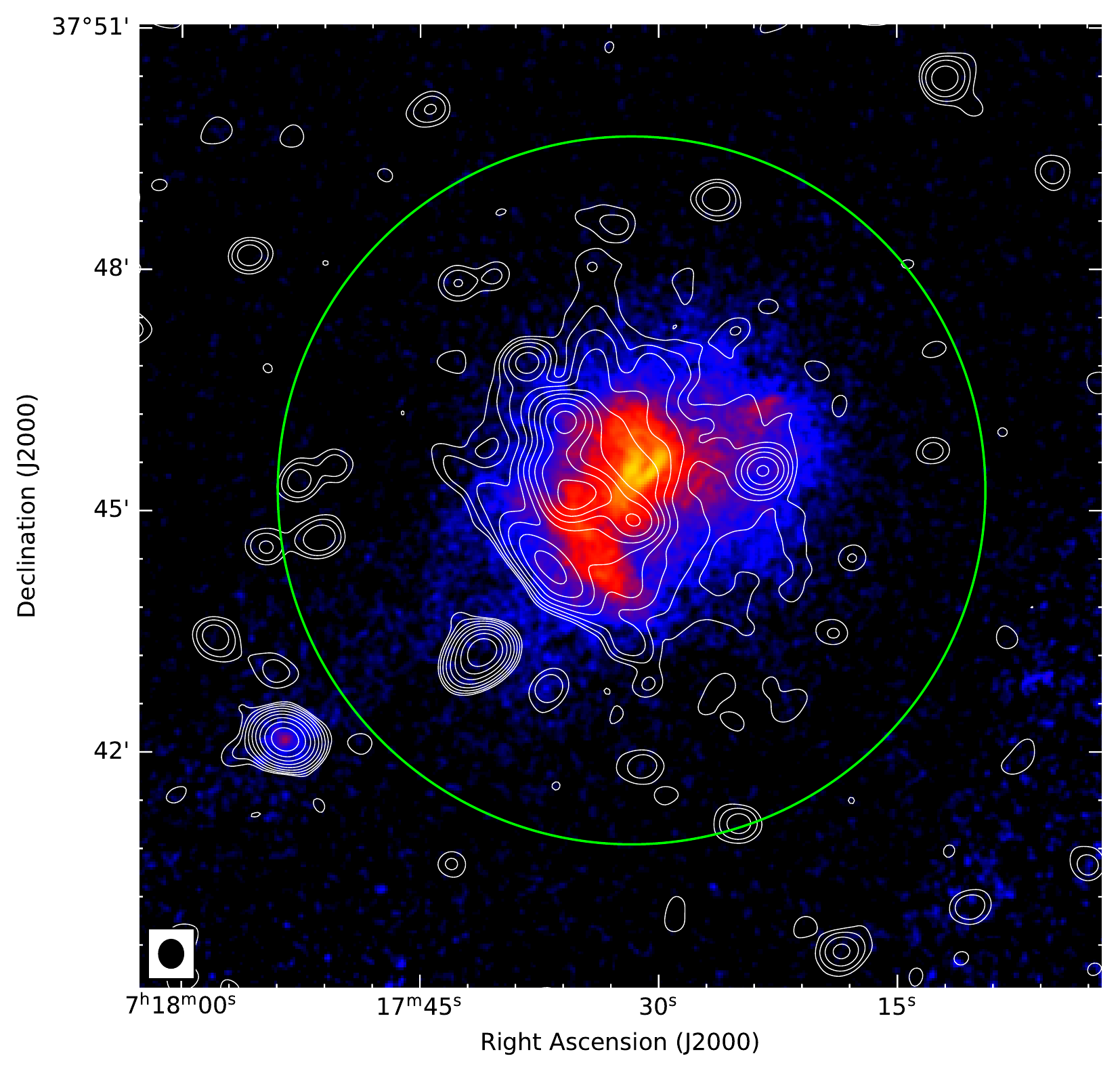}
      \vspace{-0.3cm}
    \caption{ Low resolution ($20\arcsec$) LOFAR (\textit{left}) and VLA  L-band (\textit{right}) radio contours, superimposed on the \textit{Chandra} X-ray (0.5$-$2 keV) image, smoothed to a FWHM of $3\arcsec$. Contour levels are drawn at $[1,2,4,8,\dots]\,\times\,4\,\sigma_{{\rm{ rms}}}$, where $\sigma_{\rm rms}$ at 144\,MHz and 1.5\,GHz are $130\,\rm \upmu Jy\,beam
   ^{-1}$ and  $14\,\rm \upmu Jy\,beam^{-1}$ respectively. The radio halo shows a strikingly similar morphology to the ICM distribution (traced by the X-ray emission), confirming the connection between the hot gas and relativistic plasma. The emission from the radio halo at 144\,MHz, further extends over the whole region of detected X-ray emission. The green circle shows $\rm R_{500}$ for the cluster.}
      \label{fig1a}
\end{figure*}

In Fig.\,\ref{fig1a}, we compare the X-ray morphology of the cluster to that of the radio emission at 1.5\,GHz and 144\,MHz. As mentioned in Sect.\,\ref{results}, we also detect radio emission to the northern and western parts of the cluster. The X-ray and the radio morphologies are strikingly similar, indicating a connection between the thermal gas and relativistic plasma in this system. At 1.5\,GHz, the halo emission appears to extend further to the west, roughly tracing the X-ray morphology. At 144\,MHz, the emission from the radio halo also extends to the north, encompassing the entirety of the X-ray emission region. Moreover, the emission from the halo at 144\,MHz, further extends over the whole region of detected X-ray emission.

%% I_R-I_X

\subsection{X-ray and radio surface brightness correlation}
\label{IRX}
To investigate any possible correlation between the radio and X-ray brightness, we make use of radio images created at $8\arcsec$ resolution. The \textit{Chandra} point-source-subtracted X-ray image was smoothed with a Gaussian of $3\arcsec$ full width at half maximum (FWHM). We construct a grid covering the entire halo region, including the filaments F1 and F2.  A recent study by \cite{Ignesti2020} showed that for poorly sampled sources (<30 cells), the position of the grid can affect the outcome of such an analysis. However, due  to the numerous data points and large pivot in radio and X-ray brightness spanned by data, our results are not affected by changes in the position of the grid. 

With the high resolution and high sensitivity of the available LOFAR, VLA, and \textit{Chandra} X-ray data, we can perform a detailed investigation of the point-to-point radio/X-ray correlation at 144\,MHz, 1.5\,GHz and 3\,GHz. To retain good SNR, we include only those areas where the radio surface brightness exceeds the $3\sigma$ level.

\begin{figure*}[!thbp]
\centering
\includegraphics[width=1\textwidth]{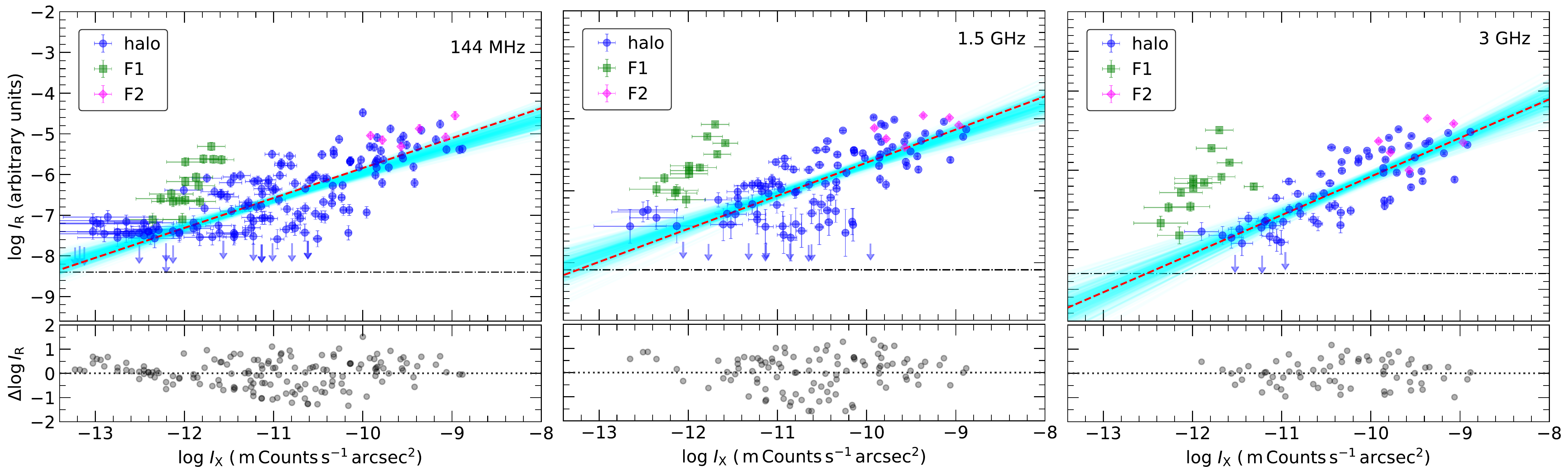}
\vspace{-0.6cm}
\caption{$I_{\rm R}{-}I_{\rm X}$ relation of the halo in MACS\,J0717.5$+$3745, extracted in square boxes with of $8\arcsec$ (about 50 kpc). {\tt Linmix} best-fit relations are indicated by red dashed lines. The upper limits represent cells with data points below $2\sigma$ radio noise level. The horizontal black dash-dotted lines indicate the $1\sigma$ in the radio maps. The best-fit parameters are obtained for the halo emission (blue dots). The cyan lines show  samples from the posterior distribution. The lower panel shows the residuals of log$\, I_{\rm R}$ and log$\,I_{\rm X}$ with respect  to the {\tt Linmix} best fit line. For the halo, the radio brightness correlates well with the X-ray at all three frequencies. The best-fitting slopes at 144\,MHz, 1.5\,GHz and 3\,GHz are $0.67\pm0.05$, $0.81\pm0.09$, and $0.98\pm0.09$, respectively. The strong dependence of the slope on frequency suggests that the halo shows a spectral steepening towards high frequencies. Displayed errors in the radio flux density measurements include the statistical and systematic uncertainties.}
\label{xrayhalo}
\end{figure*}

Fig.\,\ref{xrayhalo} shows the point-to-point comparison between the X-ray and radio brightness in log-log scale. Despite a complex distribution of thermal and non-thermal emission components in MACS\,J0717.5$+$3745, Fig.\,\ref{xrayhalo} shows a clear positive correlation: higher radio brightness is associated with higher X-ray brightness. A number of previous radio halo studies have reported a relationship between the radio and X-ray brightness \citep[e.g.,][]{Govoni2001a,Govoni2001b,Shimwell2014,Rajpurohit2018,Hoang2019,Cova2019,Xie2020,Botteon2020}. This relationship is generally described by a power law of the form:
\begin{equation}
\rm {log}\, \it I_{\rm R}=a+ b\,\rm {log}\,\it I_{\rm X},
\end{equation}
where a slope of $b=1$ suggests a linear relation, and $b<1$ (sub-linear) indicates that radio brightness increases more slowly than X-ray brightness or vice versa (if $b>1$).

To quantify the strength of any possible correlation, and to determine the best-fitting parameters to the observed data of the $I_{\rm R}{-}I_{\rm X}$ relations, we adopt the {\tt Linmix}\footnote{\url{https://linmix.readthedocs.io/en/latest/src/linmix.html}} package \citep{Kelly2007}. {\tt Linmix} performs a Bayesian linear regression and accounts for measurement uncertainties on both variables, intrinsic scatter, and upper limits (non-detections in the y-variable). It uses a Markov Chain Monte Carlo (MCMC) method and the output parameters are randomly drawn from the posterior distributions. The correlation strength was measured by using the Spearman and Pearson correlation coefficients. 

\setlength{\tabcolsep}{3pt}
\begin{table}
\caption{{\tt Linmix} fitting slopes and Spearman ($r_{s}$) and Pearson ($r_{p}$) correlation coefficients of the data for Fig.\,\ref{xrayhalo}.}
\centering
\begin{threeparttable} 
\begin{tabular}{ c c c c  c c}
 \hline  \hline  
$\nu$& $b$& $b_{\rm \,upper \,limits}$ &$\sigma_{\rm int}$ &$r_{s}$ &$r_{p}$ \\ 
\hline  
144\,MHz& $0.67\pm0.05$&$0.69\pm0.06$&$0.32\pm0.04$&0.73&0.75 \\ 
1.5\,GHz &$0.81\pm0.09$&$0.84\pm0.09$&$0.36\pm0.08$&0.70&0.69 \\ 
3.0\,GHz& $0.98\pm0.09$&$0.96\pm0.09$&$0.29\pm0.05$&0.81&0.80 \\
\hline 
\end{tabular}
\end{threeparttable} 
\label{fit}   
\end{table}

In Table\,\ref{fit}, we summarize the best-fit slopes and corresponding correlation coefficients for each radio frequency considered here. We find that the $I_{\rm R}$ and $I_{\rm X}$ are strongly correlated at 3 GHz. At this frequency, the slope is very close to linear, with $b_{3\,\rm GHz}=0.98\pm0.09$, and shows low intrinsic scatter, $\sigma_{\rm int, 3\,\rm GHz}=0.29\pm0.05$. This implies that the relativistic particles and the magnetic field are connected to the thermal plasma \citep[e.g.,][]{Govoni2001a}. Toward lower frequencies, we find sub-linear relationships with similar scatter. At 1.5\,GHz, we find $b_{1.5\,\rm GHz}=0.81\pm0.09$ and $\sigma_{\rm int, 1.5\,\rm GHz} = 0.36\pm0.08$, whereas at  144\,MHz, $b_{144\,\rm MHz}=0.67\pm0.05$ and $\sigma_{\rm int, 144\,\rm MHz} = 0.32\pm0.04$.

Across the frequency range 144\,MHz to 3\,GHz, we find that scatter on the relation is $\sim0.29-0.36$. This scatter likely arises as a result of both measurement uncertainties and intrinsic dispersion due the physical properties of the halo, for example inhomogeneous turbulence and fluctuations of the magnetic field and density of relativistic electrons. We also studied the impact of setting a threshold in $\sigma$ on the radio measurements of the $I_{\rm R}{-}I_{\rm X}$ correlation. We apply a threshold of $2\sigma$ to the radio image to check if this introduces a bias in the correlation and intrinsic scatter. The data points below the $2\sigma$ level are indicated with an upper limit in Fig.\,\ref{xrayhalo}. The best-fit slope $b$ and intrinsic scatter $\sigma_{\rm{int}}$ determined using this lower threshold of $2\sigma$ are consistent with the values obtained using a $3\sigma$ cutoff; these parameters are also shown in Table\,\ref{fit}. 

In the literature, point-to-point analysis of radio halos has been investigated mostly at a single frequency. The one exception is the halo in Abell 520, where \citet{Hoang2019} investigated the $I_{\rm R}{-}I_{\rm X}$ relation at three different frequencies. They found a tentative trend only for the central regions of the halo. However, the correlation slope in Abell 520 is broadly constant as a function of frequency: \citeauthor{Hoang2019} found $b_{145\,\rm{MHz}}=0.34\pm0.11$, $b_{323\,\rm{MHz}}=0.27\pm0.10$, and $b_{1.5\,\rm{GHz}}=0.25\pm0.09$ at 145\,MHz, 323\,MHz, and 1.5\,GHz, respectively. In contrast, for the halo in MACS\,J0717.5$+$3745, we find that the $I_{\rm R}{-}I_{\rm X}$ correlation depends significantly on the observing frequency. The correlation steepens towards high frequency, from $b = 0.67$ at 144\,MHz to $b = 0.98$ at 3\,GHz. This implies that the halo shows a spectral steepening towards high frequencies, as also found from direct spectral index and curvature analysis in Sect.\,\ref{discussion}.

Since the halo in MACS\,J0717.5$+$3745 is more extended towards lower frequencies, the regions selected for the $I_{\rm R}{-}I_{\rm X}$ correlation in Fig.\,\ref{xrayhalo} are different at different frequencies. This implies that we are sampling different regions at 144\,MHz, 1.5\,GHz, and 3\,GHz and this may affect the correlation slopes. To investigate this, we performed the same fit again, this time selecting only those regions that are detected above the $3\sigma$ level at all three radio frequencies. From this fit, we obtained the correlation slopes of $b_{144\,\rm{MHz}}=0.62\pm0.08$, $b_{1.5\,\rm{GHz}}=0.73\pm0.08$, and $b_{3\,\rm{GHz}}=0.98\pm0.09$. These values are slightly flatter than those reported in Table\,\ref{fit} (which were obtained using different regions at different frequencies) although they still show the same trend---the slope becomes more sublinear toward lower frequencies.

Fig.\,\ref{xrayhalo} also shows the $I_{\rm R}{-}I_{\rm X}$ relation for the filaments F1 and F2, indicated respectively by green and magenta points. F1 shows clearly distinct behavior at all three frequencies, suggesting it may not be associated with the halo emission. In the radio vs X-ray correction F2 does not appear distinct may be due to the superposition of the halo emission and F2. Both F1 and F2 are also characterized by a high degree of polarization ($>15\%$; Rajpurohit et al. to be submitted). These evidence suggests that these two structures are not associated with the halo emission and thus have a different origin like shock fronts.

\setlength{\tabcolsep}{13pt}
\begin{table}
\caption{{\tt Linmix} best fitting slopes ($b$) and Spearman ($r_{s}$) and Pearson ($r_{p}$) correlation coefficients of the data for the left panel of Fig.\,\ref{Xray_index} ($I_{\rm X}-\alpha$ correlation).}
\centering
\begin{threeparttable} 
\begin{tabular}{ c c c c  c c}
 \hline  \hline  
& H1+H2&  H1 &H2 \\ 
\hline  
$b$& $-0.07$& $-0.09$& $-0.20$ \\
$r_{s}$& $-0.36$& $-0.82$& $-0.76$\\ 
$r_{p}$& $-0.52$& $-0.67$ &$-0.62$\\
\hline 
\end{tabular}
\end{threeparttable} 
\label{fit_index}   
\end{table}

%% alpha-I_X

\subsection{X-ray brightness vs spectral index}
We also studied the point-point distribution of halo spectral index with thermal gas. To extract the X-ray surface brightness and spectral indices, we use the same grid regions mentioned in Sect.\,\ref{index}. As the halo is more extended towards low frequencies, we extract spectral indices between 144\,MHz to 1.5\,GHz.

\begin{figure}[!thbp]
    \centering
       \includegraphics[width=0.49\textwidth]{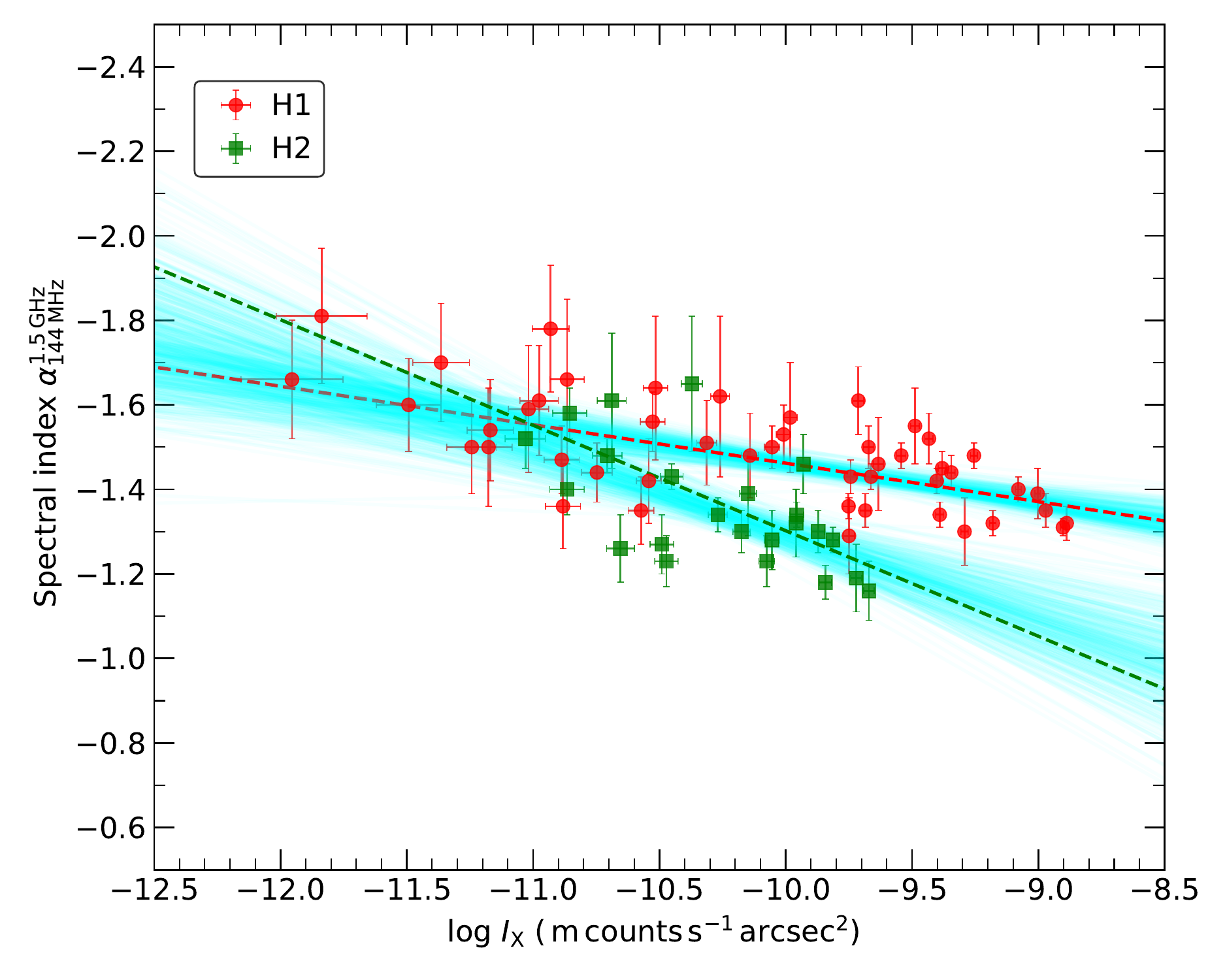}
       \vspace{-0.8cm}
           \caption{ X-ray surface brightness versus spectral index  of the radio halo emission. The Bayesian linear regression is performed for data points with at least a $3\sigma$ radio noise level. The red and green dashed lines correspond to the best-fit obtained separately for the H1 and H2 region of the halo. The anti-correlation between these two quantities is evident, indicating that the spectral index is flatter at high X-ray brightness and tends to get  steeper in low X-ray brightness regions.} 
      \label{Xray_index}
\end{figure}

The results are shown in Fig.\,\ref{Xray_index}. There is evidence of an anti-correlation between the radio spectral index and the X-ray brightness: the spectral index is flatter at high X-ray brightness and steeper at low X-ray brightness. Additionally, the two halo regions (H1 and H2) appear to show markedly different behavior in the $\alpha{-}I_{\rm R}$ plot. To check the significance of the correlation, we fit the data assuming a relation of the form:

\begin{equation}
    \alpha=a+b\,\rm {log}\,\it I_{\rm X}.   
\end{equation}

We again performed a linear regression for the halo regions above $3\sigma$ level using ${\tt Limix}$. As we are principally interested in studying the halo, we excluded the polarized regions embedded in the halo (the cyan boxes in Fig.\,\ref{halo_regions}). The resultant best-fit parameters are summarized in Table\,\ref{fit_index}. These results hint at an anti-correlation between the spectral index and X-ray surface brightness, which becomes much stronger when data points from H1 and H2 regions are fitted separately. Recently, \cite{Botteon2020} studied the $\alpha{-}I_{\rm X}$ relation for the halo in Abell 2255. They found a positive correlation and mild trend between the spectral index and X-ray surface brightness. In contrast, the halo in MACS\,J0717.5$+$3745 shows a strong negative trend. \cite{Botteon2020} also studied possible relations between spectral index of the synchrotron emission and thermodynamical properties of the ICM (namely, temperature, pseudo-pressure, and pseudo-entropy). We do not report this kind of analysis for MACS\,J0717.5$+$3745 because the low effective area of Chandra above 5 keV makes very uncertain the measurement of high temperatures, such as those reported in MACS\,J0717.5$+$3745 \citep[up to and beyond 20 keV;][]{vanWeeren2017b}, making unfeasible the study of these correlations.

%%%%%%%%%%%%%%%%%%%%%%%%%%%%%%%%%%%%%%%%%%%%%%%%%%%%%%%%%%%%%%%%%%%%%%%%%%%%%%%%%%%%
\section{Discussion}
\label{discussion1}
%%%%%%%%%%%%%%%%%%%%%%%%%%%%%%%%%%%%%%%%%%%%%%%%%%%%%%%%%%%%%%%%%%%%%%%%%%%%%%%%%%%%

Several key aspects of turbulent models and of halo properties remain poorly understood \citep[see e.g.,][for reviews]{Brunetti2014,vanWeeren2019}. One aspect is that the properties of radio halo synchrotron spectra are still poorly known, both observationally and theoretically. Radio halos show a fairly broad range of spectral indices, $-2 \lesssim \alpha \lesssim -1$ \citep{Feretti2012,vanWeeren2019}.  

The spectral analysis reveal that the  MACS\,J0717.5$+$3745 halo shows a steep spectrum ($\alpha_{144\,\text{MHz}}^{1.5\,\text{GHz}}-1.39\pm0.04$) and a steepening ($\alpha_{1.5\,\text{GHz}}^{5.5\,\text{GHz}}=-1.93\pm0.04$) above 1.5\,GHz. Moreover, we find clear evidence of spectral steepening and increasing curvature in the outermost regions of the halo.  To our knowledge, curved spectra have only been measured in two radio halos: the Coma \citep{Thierbach2003} and Abell S1063 \citep{Xie2020} clusters. On the other hand there are also few other halos with adequate datasets covering a sufficient frequency range, for example radio halos in the Toothbrush cluster \citep{Rajpurohit2019}, Sausage cluster \citep{Hoang2017,Gennaro2018}, the Bullet cluster \citep{Shimwell2014}, and A2744 \citep{Pearce2017}, where the spectra are compatible with power-laws.

In reacceleration models a high energy break in the spectrum of electrons is generated as a result of the balance between the rate of reacceleration and that of synchrotron and IC losses \citep[e.g.,][for a review]{Brunetti2014}. The spectra of halos reflect the presence of this break and are predicted to be sensitive to, both, the underlying acceleration mechanism and the physical conditions in the ICM. Homogeneous reacceleration models predict curved spectra, with a spectral steepening occurring at frequencies larger than few times the critical frequency of the highest-energy electrons reaccelerated by the mechanism \citep[e.g.,][]{Cassano2005}  $\nu_s \sim \xi \nu_c$, where $\nu_c$ is the critical frequency; $\xi\sim$ 6-8 was estimated by \citet{Cassano2012} (their Fig. 2). On the other hand, if the magnetic field, the acceleration rate and the timescale of turbulence generation (and decay) vary across the emitting volume and along the line of sight, the observed spectrum will be a blend of different components. As a result, the spectrum can  get stretched in frequency, with the spectral curvature becoming progressively less evident in the case of less homogeneous conditions \citep[see simulations by][]{Donnert2013,ZuHone2011}.

Nevertheless, even in these conditions the presence of a maximum energy in the spectrum of electrons generates a steepening in the synchrotron spectrum. The steepening is expected few times above the critical frequency  of the electrons accelerated in the regions where the acceleration efficiency is larger and interacting with a magnetic field $B \sim B_{\rm cmb}/\sqrt{3}$ \citep[e.g.,][]{Brunetti2016}:

\begin{equation}
{\frac{\nu_s}{{\rm{GHz}}}} \sim \left( \frac{420}{\tau_{\rm acc}/{\rm{Myr}}} \right)^2 (1+z)^{-7},
\label{eq:maxnu}
\end{equation}
where $\tau_{\rm acc}$ is the minimum acceleration time in the emitting volume.

MACS\,J0717.5$+$3745 is at higher redshift than the majority of radio halos studied to-date. At higher redshift, IC losses limit the maximum energy of electrons that are reaccelerated increasing the chance to detect a spectral steepening. We detect a steepening above 1.5\,GHz, implying that the minimum acceleration time in the emitting volume is $\tau_{\rm acc} \sim 80$ Myr (using Eq.,\ref{eq:maxnu}), provided that the fluctuations of magnetic field in the emitting volume reach values $B \sim B_{\rm cmb}/\sqrt{3} \sim 4.5 \upmu$G.

For the halo in MACS\,J0717.5$+$3745, we find a mean scatter of 0.28 and 0.53 around the mean spectral index between 144$-$1500\,MHz and 1.5$-$3\,GHz, respectively. Such a large scatter is likely due to the strong IC losses at the high redshift of MACS\,J0717.5$+$3745 that quench the acceleration process where turbulence is lower. In a large fraction of the emitting volume, under these conditions the fraction of energy available for acceleration that is dissipated into synchrotron radiation scales with $B^{2}/B_{\rm IC}^{2}$ implying fast changes of the emissivity with $B^{2}$ fluctuations. Moreover, in a steep spectral index regime the changes of $\tau_{\rm acc}$ (Eq.\,\ref{eq:maxnu}) induce strong variations of the emissivity and spectrum. The measured scatter implies that turbulence varies when averaged on about 50 -100 kpc cells and integrated along the line of sight. 

The mean scatter of the spectral index shows a moderate decrement from the analysis performed with 50 kpc to 90 kpc box size. On one hand this decline is expected  because a larger number of independent regions in the  halo are intercepted by adopting larger cells of the grid. However if these regions would be much smaller than the cell size the scatter should decline with the inverse of the cell size \citep[e.g.,][]{Botteon2020}. The fact that the scatter declines less rapidly suggests that size of the regions is of the same order of that of the cells. This also suggests that turbulent injection scales cannot be less than 50-100\,kpc otherwise we would see a smoother distribution.

Similar to the case of the Coma halo, we also detect a spectral steepening toward the halo periphery in MACS\,J0717.5$+$3745. Following \cite{Brunetti2001}, the steepening can be interpreted as an effect of the decline of the magnetic field strength with radius, being $\nu_s \propto B$. This explanation also holds in the case of more complicated models: if the magnetic field has a broad distribution of values in the emitting volume, the decline of the mean value of the magnetic field with radius also implies that a progressively smaller fraction of the volume is filled with magnetic field reaching values $B \sim B_{\rm cmb}/\sqrt{3}$, which implies a gradual steepening of the spectrum with distance.

Interestingly, the limits on the turbulent acceleration timescale that we obtained from the spectral steepening at higher frequencies can also be used to constrain the energy density and scales of turbulence. This of course depends on the particular reacceleration model that is adopted, as different scenarios considered for the non-linear interaction between turbulent fluctuations and particles have implications on the fraction of the turbulent energy flux that is effectively damped into particle acceleration.

\cite{Brunetti2016} proposed a stochastic reacceleration model, based on the scattering of particles by magnetic field lines diffusing into super-Alfvenic incompressible turbulence. In this model, the observed steepening frequency ($\nu_s$; above which the spectrum steepens) and turbulent Mach number $(M_{\rm t})$ are connected via:
\begin{equation}
M_{\rm t} \sim 0.2 \frac{ 2000\,{\rm{km}}\,{\rm{s}}^{-1}}{c_s} \left(
\frac{\nu_{\rm s}\,(\rm GHz)}{1.5} \right)^{\sfrac{1}{6}}
\left(
\frac{ L}{10^2 \rm \,kpc} \right)^{\sfrac{1}{2}} (1+z)^{\sfrac{11}{6}},
\label{eq:Mach}
\end{equation}
where $c_s$ is the speed of sound in the medium  and $L$ is turbulent injection scale. In the case of MACSJ\,0717.5$+$3745, Eq.\,\ref{eq:Mach} implies a ratio of turbulent (solenoidal) and thermal energy densities 
$\epsilon_t/\epsilon_{\rm ICM} \sim 1/3 \Gamma M_{\rm t}^2 \sim 0.16$, considering injection scales $L\sim 100$ kpc and an adiabatic index $\Gamma =5/3$. Similar constraints are obtained by considering different acceleration mechanisms. For example, if we assume transit-time-damping (TTD) that is a reference model for radio halos \citep[e.g.,][]{Brunetti2007,Miniati2015} and is based on the TTD resonance between particles and fast modes we obtain a ratio of turbulent (compressive) and thermal energy densities $\epsilon_t/\epsilon_{\rm ICM} \sim 0.1$, considering injection scales $\sim 100$ kpc. The flux of kinetic energy from the turbulent motions required above appears to be in line with what is measured in cosmological simulations of turbulent motions in the ICM \citep[e.g.,][]{2012A&A...544A.103V,2017MNRAS.464..210V}.

Despite the complex morphology of the halo, we find the radio brightness correlates strongly with the X-ray brightness. The radio and X-ray brightness correlation slopes steepen at higher  frequency (from $b_{144\,\text{MHz}}=0.67\pm0.05$ to $b_{3\,\text{GHz}}=0.98\pm0.09$). Previous studies of the $I_{\rm R}{-}I_{\rm X}$ point-to-point correlation have shown that the best-fit slopes vary over a broad range of values \citep[e.g.,][]{Govoni2001a,Govoni2001b,Rajpurohit2018,Ignesti2020,Botteon2020} and are mainly flatter than 1 (i.e., sub-linear). For a purely hadronic origin scenario of giant radio halos, a significantly steeper slope is expected for the relation linking the radio and X-ray brightness \citep{Govoni2001a,Pfrommer2008}. 

The sub-linear $I_{\rm R}{-}I_{\rm X}$ correlation slopes in the case of MACS\,J0717.5$+$3745 halo indicates that the ratio of the radio and X-ray surface brightness increases where the steepening occurs (mainly in the outermost regions). Since the SZ decrement is less in the external regions (due to the decreased pressure and path length through the cluster, assuming roughly spheroidal geometry), this also suggest that SZ effect cannot be responsible for the observed steepening. The radio-and X-ray surface brightness correlation slopes steepen at higher radio frequency (from $b_{144\,\text{MHz}}=0.67\pm0.05$ to $b_{3.0\,\text{GHz}}=0.98\pm0.08$), In addition, we find a significant anti-correlation between the X-ray surface brightness and the spectral index.Both pieces of evidence further support a spectral steepening in the external regions.

%%%%%%%%%%%%%%%%%%%%%%%%%%%%%%%%%%%%%%%%%%%%%%%%%%%%%%%%%%%%%%%%%%%%%%%%%%%%%%%%%%%%
\section{Summary and conclusions}
\label{summary}
%%%%%%%%%%%%%%%%%%%%%%%%%%%%%%%%%%%%%%%%%%%%%%%%%%%%%%%%%%%%%%%%%%%%%%%%%%%%%%%%%%%%

In this work, we have presented new 144\,MHz LOFAR HBA radio observation of the galaxy cluster MACS\,J0717.5$+$3745. These observations were combined with existing uGMRT (330$-$850\,MHz) and VLA  (1$-$6.5\,GHz) observations in order to carry out a detailed spectral analysis of the halo and its connection with the ICM. Our deep LOFAR observations, processed using the state-of-the-art LoTSS pipeline, yielded sensitive, high-resolution images which have allowed us to recover faint emission that was not detected previously at this frequency. We summarize the overall results as follows:

\begin{enumerate} 
\item{} The new LOFAR images recover both the halo and the relic, previously reported at the same frequency. At 144\,MHz, the halo emission is more extended to the north and northwest of the cluster center. Our data reveal that the halo becomes significantly more extended moving toward lower frequencies from 5.5\,GHz to 144\,MHz, suggesting steepening in outermost regions. \\

\item{} The integrated spectrum of the halo shows a clear spectral break above 1.5\,GHz, resulting in a spectral steepening toward higher frequencies. The integrated spectral index between 144\,MHz and 1.5\,GHz is $-1.39\pm0.04$, steepening to $-1.93\pm0.04$ between 1.5 and 3.0\,GHz. The high frequency steepening cannot be explained solely by a decrement due to the SZ effect.  \\

\item{} The spatially-resolved spectral index maps show a significant spectral variations. We find a mean scatter of $0.28$ and $0.53$ around the mean spectral index from 144\,MHz-1.5\,GHz and 1.5\,GHz - 3.0\,GHz, respectively. Such a strong scattering in the spectral index may be due to the strong inverse Compton losses at the high redshift of MACS\,J0717.5$+$3745. Our study of spectral variations also suggests that the turbulent injection scales cannot be much less than $50-100$\,kpc. \\

\item{} The distributions of spectral index show clear evidence of spectral steepening in the external regions of the halo. In the innermost regions of the halo, the spectral index remains largely constant at around $\alpha=-1.35$. Conversely, moving toward the outermost regions, the spectrum steepens significantly, reaching values of of about $-3.0$. \\

\item{}  The spatially-resolved spectral curvature maps show significant curvature variations across the halo. The central regions of the halo shows little-to-no curvature, whereas curvature increases significantly in the outermost regions of the halo. Color-color analysis of regions within the radio halo also show spatial variations in the curvature. \\

\item{} Despite the complex distribution of thermal and non-thermal emission components, the morphology of the radio halo emission is  similar to the X-ray emitting gas, confirming the connection between the hot gas and relativistic plasma in this system. The radio brightness correlates strongly with the X-ray brightness at all observed frequencies.  The filaments F1 appear distinct in the radio vs. X-ray surface brightness correlation, and is likely not associated with the halo emission.\\

\item{}  The radio vs. X-ray surface brightness correlation steepens toward higher radio frequencies, namely from $0.67\pm0.05$ (at 144\,MHz) to  $0.98\pm0.09$ (at 3\,GHz). This provides further evidence that the spectral index steepens in the peripheral regions of the halo.\\

\item{} An anti-correlation is found between the radio spectral index and the X-ray surface brightness, implying that the spectral index is flatter at high X-ray brightness and steeper at low X-ray brightness. This is consistent with the fact that there is a radial steepening across the halo.

\end{enumerate} 

For the halo in MACS\,J0717.5$+$3745, the compelling evidence for radial steepening, the existence of a spectral break above 1.5\,GHz, the strong dependence of the $I_{\rm R}{-}I_{\rm X}$ correlation on frequency, and the anti-correlation between the spectral index against the X-ray brightness can be interpreted in the context of turbulent reacceleration models.  In this scenario we estimate that the minimum acceleration time in the emitting volume that is necessary to match the spectrum of the halo considering its redshift is about 80 Myr and that the turbulent energy density in the regions where reacceleration is stronger is $\sim 10\%$ of the thermal ICM. This amount of turbulence appears a natural consequence of the very active dynamics of the system.

%%%%%%%%%%%%%%%%%%%%%%%%%%%%%%%%%%%%%%%%%%%%%%%%%%%%%%%%%%%%%%%%%%%%%%%%%%%%%%%%%%%%%
\begin{acknowledgements}
%%%%%%%%%%%%%%%%%%%%%%%%%%%%%%%%%%%%%%%%%%%%%%%%%%%%%%%%%%%%%%%%%%%%%%%%%%%%%%%%%%%%

We thank the anonymous reviewer for the constructive feedback. KR and FV acknowledges financial support from the ERC Starting Grant "MAGCOW", no. 714196. GB and RC acknowledge support from INAF mainstream program ``Galaxy clusters science with LOFAR". AB, CJR, EB, and MB acknowledges support from the ERC through the grant ERC-Stg DRANOEL n. 714245. RJvW and AB acknowledges support from the VIDI research program with project number 639.042.729, which is financed by the Netherlands Organization for Scientific Research (NWO). W.F.  acknowledges support from the Smithsonian Institution and the Chandra High Resolution Camera Project through NASA contract NAS8-03060. GDG acknowledges support from the ERC Starting Grant ClusterWeb 804208. LOFAR \citep{Haarlem2013} is the Low Frequency Array designed and constructed by ASTRON. It has observing, data processing, and data storage facilities in several countries, which are owned by various parties (each with their own funding sources), and that are collectively operated by the ILT foundation under a joint scientific policy. The ILT resources have benefited from the following recent major funding sources: CNRS-INSU, Observatoire de Paris and Universit\'{e} d'Orl\'{e}ans, France; BMBF, MIWF-NRW, MPG, Germany; Science Foundation Ireland (SFI), Department of Business, Enterprise and Innovation (DBEI), Ireland; NWO, The Netherlands; The Science and Technology Facilities Council, UK; Ministry of Science and Higher Education, Poland; The Istituto Nazionale di Astrofisica (INAF), Italy. This research made use of the LOFAR-UK computing facility located at the University of Hertfordshire and supported by STFC [ST/P000096/1], and of the LOFAR-IT computing infrastructure supported and operated by INAF, and by the Physics Dept. of Turin University (under the agreement with Consorzio Interuniversitario per la Fisica Spaziale) at the C3S Supercomputing Centre, Italy. This research also made use of the computing facility at Th\"uringer Landessternwarte (TLS), Germany. This research made use of computer facility  on the HPC resources at the Physical Research Laboratory (PRL), India.  The scientific results reported in this article are based in part on observations made by the \textit{Chandra} X-ray Observatory and published previously in \citet{vanWeeren2017b}. We thank the staff of the GMRT that made these observations possible. The GMRT is run by the National Centre for Radio Astrophysics (NCRA) of the Tata Institute of Fundamental Research (TIFR).
\end{acknowledgements}

\bibliographystyle{aa}

\bibliography{ref.bib}

\end{document}